%% file: arxiv-template.tex
\documentclass[times,review,nopreprintline]{elsarticle}

\usepackage{caption}
\usepackage{subcaption}
\usepackage{wrapfig}
\usepackage{float}

\usepackage{arxiv}
\usepackage{framed,multirow}

\usepackage{amssymb}
\usepackage{latexsym}

\usepackage{url}
\usepackage{xcolor}
\usepackage{soul}

\usepackage{graphicx}
\graphicspath{{figures/}}

\usepackage{amsmath}

\usepackage{afterpage}

\usepackage{tikz}
\usetikzlibrary{shapes.arrows}

\definecolor{newcolor}{rgb}{.8,.349,.1}

\newcommand{\transcolor}{0} 
\newcommand{\reva}[1]{{\sethlcolor{yellow!\transcolor}\hl{#1}}}
\newcommand{\revb}[1]{{\sethlcolor{cyan!\transcolor}\hl{#1}}}
\newcommand{\revc}[1]{{\sethlcolor{green!\transcolor}\hl{#1}}}

\journal{Journal of Computational Physics}

\makeatletter
\def\input@path{{Inputs}}
\makeatother
\graphicspath{{Inputs/}
              {Figures/}
                {Figures/Entropy_Plots/}
                  {Figures/Entropy_Plots/BW/}
                  {Figures/Entropy_Plots/OT/}
                  {Figures/Entropy_Plots/SBdens/}
                  {Figures/Entropy_Plots/Sedov/}
                {Figures/Failures_Corr/}
                  {Figures/Failures_Corr/BW/}
                  {Figures/Failures_Corr/SBdens/}
                {Figures/IC/}
                {Figures/Logos/}
                {Figures/Motivation/}
                {Figures/Recovery/}
                  {Figures/Recovery/BW/}
                  {Figures/Recovery/OT/}
                  {Figures/Recovery/SB/}
                  {Figures/Recovery/Sedov/}}

\begin{document}

\verso{Gina Vasey \textit{etal}}

\begin{frontmatter}

\title{Influence of initial conditions on data-driven model identification and information entropy for ideal mhd problems}

\author[1]{Gina Vasey \fnref{fn1}}
\fntext[fn1]{Corresponding author email: vaseygin@msu.edu}

\author[2]{Daniel Messenger}

\author[2]{David Bortz}

\author[1,3]{Andrew Christlieb}

\author[1,4]{Brian O'Shea}

\address[1]{Computational Mathematics, Science and Engineering, Michigan State University, East Lansing, MI 48824, USA}
\address[2]{Applied Mathematics, University of Colorado Boulder, Boulder, CO 80309, USA}
\address[3]{Mathematics, Michigan State University, East Lansing, MI 48824, USA}
\address[4]{Physics and Astronomy, Michigan State University, East Lansing, MI 48824, USA}

\received{6 December 2023}

\begin{abstract}
Data-driven methods of model identification are able to discern governing dynamics of a system from data. Such methods are well suited to help us learn about systems with unpredictable evolution or systems with ambiguous governing dynamics given our current understanding. Many plasma problems of interest fall into these categories as there are a wide range of models that exist, however each model is only useful in a certain regime and often limited by computational complexity. To ensure data-driven methods align with theory, they must be consistent and predictable when acting on data whose governing dynamics are known. Weak Sparse Identification of Nonlinear Dynamics (WSINDy) is a recently developed data-driven method that has shown promise in learning governing dynamics from data with high noise levels \cite{wsindy_pde}. This work examines how WSINDy acts on ideal MHD test problems as the initial conditions are varied and specifies limiting requirements for successful equation identification. It is hard to recover the governing dynamics from data that emphasize a single dominant behavior. In these low information cases, Shannon information entropy is able to pick up on the redundancies in the data that affect recoverability.
\end{abstract}

\begin{keyword}


\MSC[2020] 35Q60\sep 35R30\sep 76W05\sep 85A30

\KWD weak sparse identification of nonlinear dynamics\sep data-driven\sep ideal MHD\sep shannon information entropy
\end{keyword}

\end{frontmatter}


\section{Introduction}
\input{Intro.tex}

\section{Methods}\label{sec:methods}
\input{Methods.tex}

\section{Experiment}\label{sec:exp}
\input{Exp.tex}

\section{Results}\label{sec:results}
\input{Results}

\section{Discussion}\label{sec:disc}
\input{Discussion.tex}

\section{Conclusions and Future Work}\label{sec:conc}
\input{Conclusions.tex}

\section*{Data Availability Statement}
The data that support the findings of this study are available from the corresponding author upon reasonable request.





\section{Acknowledgements}
This material is based upon work supported by: the U.S. Department of Energy, Office of Science, Office of Advanced Scientific Computing Research under Award Numbers DE-SC0023164 and DE-SC0023346; the National Science Foundation under Grant No. 2008004.

\bibliographystyle{model1-num-names}
\bibliography{refs}




\end{document}

%% file: Intro.tex

\begin{figure}
    \centering
    \input{regimes_dpf}
    \caption{Approximate span of regimes of Dense Plasma Focus (DPF), annotated on plot of approximate model types and their applicable regimes \cite{SandiaReport,uri}. The x-axis represents the Knudsen Number $K_n$, a measure of plasma collisionality relative to simulation scale. The y-axis is the Charge Separation $\Lambda$, the length scale at which electromagnetic forces act. Particle simulations are well suited for less collisional plasma. Fluid-like models are most appropriate when the collisionality of the plasma increases, but extra terms that approximate particle behavior are needed. Fluid models are well suited for highly collisional plasma. Each of these zones encompass a family of models, described in full detail in \cite{SandiaReport}. A DPF begins in a less collisional regime around point A, requiring particle based simulations like particle-in-cell (PIC) methods, and resolves in a collisional regime with strong electromagnetic effects around point B where fluid models, specifically ideal magnetohydrodynamics (MHD), would be most appropriate.}
    \label{fig:DPF}
\end{figure}
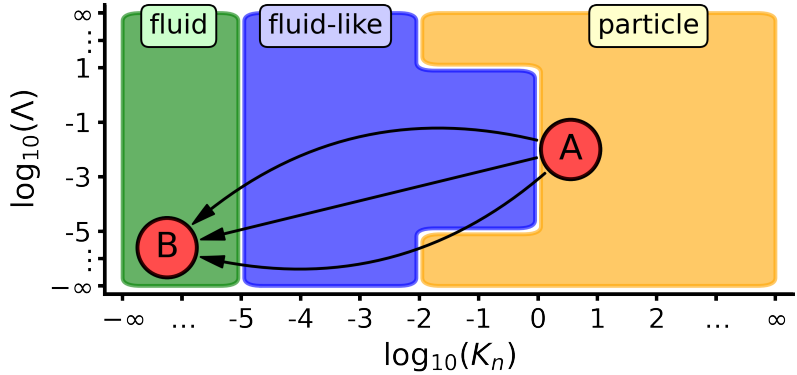

Many experimental plasma systems pass through multiple regimes that require different plasma models and therefore simulation methods to accurately capture their behavior. An estimate of the collisionality of the plasma relative to the simulation scale (the Knudsen Number $K_n$) and the length scale at which electromagnetic forces interact (the Charge Separation $\Lambda$) can be used to guide what plasma model may best fit a system as shown in Figure \ref{fig:DPF}. However, these measures are not precise and systems that evolve through a significant portion of this space pose an extra challenge. Less collisional systems have a greater $K_n$ value and are best simulated by evolving the particle phase space distribution, shown by the particle region of Figure \ref{fig:DPF}. Some codes, like Gkeyll, solve the Vlasov equations while others, like Warp-X, solve the Boltzmann equation \cite{gkeyll, warpx}. This is often achieved by particle-in-cell (PIC) codes. As the plasma becomes more collisional, PIC codes become too computationally expensive to evolve the system over the required spatial and temporal domains. In this limit fluid approximations are introduced. These fluid models are found by integrating the particle distribution in velocity space, resulting in moments of the phase space distribution to represent fluid quantities. Fluid-like models introduce extra terms beyond what are classically thought of as fluid quantities, like density and velocity, to account for particle effects that are important in moderately collisional regimes. Some of the commonly used fluid and fluid-like models are derived in full detail in \cite{SandiaReport}. A system that moves from a less collisional to more collisional region thus requires different representations as it evolves in time.

For example the Dense Plasma Focus (DPF), an experimental setup used as a neutron and X-ray source related to fusion, is generally described as passing through three main phases \cite{DPF_apps}. The initial breakdown phase is weakly collisional and electromagnetic effects act on a shorter scale, suggesting the system resides somewhere around point A in Figure \ref{fig:DPF}. In the final dense and highly collisional pinch stage with long spanning electromagnetic effects, a fluid representation would be required, approximately around Point B in Figure \ref{fig:DPF}. Note though that some interactions in the final phase still contain some particle-driven effects, meaning an ideal MHD representation would not match experiments \cite{DPF_apps}. A common solution for modeling this wide range of behaviors is to transition between different models when some condition is met. Even doing this, though, requires knowledge of what models to transition through and when this switch must occur, both of which are ambiguous, leading to inaccurate simulation results.

A better approach would be to consider the evolution of the data and have a more precise method to pick the best model at a given point in time or space. That is, have a data-driven model identification method that can pick out the best representative model that matches the behavior close enough while not being too computationally expensive. Recently developed methods are well suited to accomplish exactly such a task. Some of these methods are neural network (NN) based; however, these require large amounts of training data, which for computationally expensive PIC codes would be impractical \cite{petersen_deep_2021,long_pde-net_2019}. Another group of methods solve a sparse regression problem. The general approach is to build a \revc{candidate library} which encompass the data's behavior. Sparse regression is employed to select a subset of the \revc{candidate library} that match the data well \cite{wsindy_ode,wsindy_pde,sindy_pde,sindy_ode}. This approach does not require a large amount of training data, and identifies a differential equation given a single dataset.


These sparse regression methods have a number of partial differential equations (PDEs) that they are generally tested on, as shown in Table 2 of \cite{wsindy_pde}. These equations only involve one or two evolved quantities and, in the case of weak form based methods, are easily integrated by parts to build their weak form representations. A few plasma cases have been considered, however not in a broad manner that indicate the generalizability of such data-driven methods to other types of plasma problems \cite{alves_data-driven_2020,kaptanoglu_physics-constrained_2021}. Thus, before using such methods to transition between regimes, this paper examines how one such sparse regression method called Weak Sparse Identification of Nonlinear Dynamics (WSINDy) performs on simulations using the same underlying model, namely ideal magnetohydrodynamics (MHD), but with different initial conditions. The ideal MHD equations consist of far more evolved quantities than the PDEs classically used. Even for this one model, there are a variety of initial conditions and behaviors of interest, some of which are more challenging for these data-driven methods to handle. Different sets of initial conditions lead to different types of interactions and therefore dominant features. There are commonly used initial conditions designed to test if specific properties are maintained by simulation methods. In this paper we use some commonly used test problems to explore from which behaviors and data conditions WSINDy can recover the ideal MHD equations. \reva{The experiments completed here serve as a validation of WSINDy on well understood plasma simulation data before moving to more complex simulations or those with evolving mechanics.}

Having descriptive metrics that coordinate with model discovery methods has been shown to enhance results. For example the structural similarity index (SSI) has been used to determine whether there is significant enough change between data samples in time \cite{Abramovic_Alves_Greenwald_2022}. Others have used ensembles of recovered PDEs along with methods like the Akaike information criteria (AIC) or Bayes information criteria to select which of the candidate models best match the data \cite{Mangan_Kutz_Brunton_Proctor_2017}. No metric has been identified, however, that reflects how descriptive the behavior contained in a dataset is of the theoretically known underlying dynamics \reva{for model identification methods. Some work has been done to analyze the complexity of the identified model or an analysis on chaotic systems has been considered, but not in the context as predictive metric for model identification \mbox{\cite{Kaptanoglu_Zhang_Nicolaou_Fasel_Brunton_2023}}}. Consider a simulation where the magnetic field is constant. Even if the system is evolved using the ideal MHD equations, there is nothing going on in the magnetic field meaning the PDE for the magnetic field cannot be extracted. This is an extreme example, but there could be some low information limit that causes a similar problem. The fields must interact sufficiently for WSINDy to identify coupled terms. It would be helpful to have a metric that could accurately predict whether WSINDy can recover certain behavior from a dataset. Such a metric would provide guidance as to whether there is enough information in a dataset to learn about certain dynamics. For this we explore the Shannon information entropy metric, as it is a measure of the amount of information in a system as reflected by its probability distribution of states \cite{engelmann_towards_2022}.

In this paper we start in Section \ref{sec:methods} with a discussion of the two main methods used (WSINDy and Shannon information entropy). Section \ref{sec:exp} introduces the simulation data used and the associated WSINDy hyperparameters for each dataset. Section \ref{sec:results} provides results for each dataset while Section \ref{sec:disc} compares and contrasts the results between the datasets. Finally Section \ref{sec:conc} summarizes our findings and provides suggestions for future work.

%% file: regimes_dpf.tex
\includegraphics[width=0.65\textwidth]{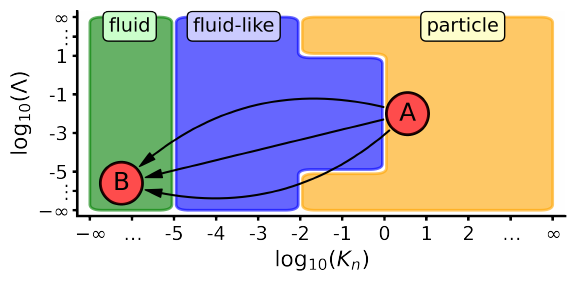}

%% file: Methods.tex

The primary methods used in this work are Weak Sparse Identification of Nonlinear Dynamics (WSINDy) and Shannon information entropy. Section \ref{sec:wsindy} describes WSINDy, a data-driven model discovery method that is used to identify equations that govern the dynamics of data. Section \ref{sec:sie} introduces Shannon information entropy, a metric that can be used to provide insight into the ability of the WSINDy algorithm to recover said dynamical equations.

\subsection{Data-Driven Model Identification with WSINDy}\label{sec:wsindy}

Weak Sparse Identification of Nonlinear Dynamics (WSINDy) is a method of data-driven model discovery that performs well on noisy simulation data \cite{wsindy_ode, wsindy_pde}. Consider a PDE in the form of Equation \ref{eq:strong} over spatial and time domains $\Omega$ and $(0,T)$ respectively.

\begin{equation}
    \partial_t u(x,t) = F(u(x,t))
    \label{eq:strong}
\end{equation}
\begin{equation}
    \partial_t u(x,t) \approx \Theta \Xi
    \label{eq:sindy}
\end{equation}

Sparse identification of Nonlinear Dynamics (SINDy) is the predecessor to WSINDy, and works directly from the strong form of the PDE in Equation \ref{eq:strong}. SINDy approximates the function $F$ by identifying as few terms as possible from a candidate library that matches the data $\partial_t u(x,t)$ well enough \cite{sindy_ode, sindy_pde}. This form is shown in Equation \ref{eq:sindy} with candidate library $\Theta$ and sparse matrix $\Xi$ which contains the learned coefficients of each term in the candidate library.

The candidate library $\Theta$ in Equation \ref{eq:sindy} consists of trial functions $(f_j(x,t))_{j \in \lbrace J \rbrace}$ that make up a guess as to what types of terms the governing PDE could consist of. Making this library too large leads to increased computational cost and can cause numerically equivalent terms to be considered simultaneously, making the sparse regression more difficult. In this work we focus on candidate libraries that contain spatial derivatives of monomial terms in the state variables, identified by a maximum monomial order and maximum derivative order.

\revb{The following sections go into more detail of the various components of the WSINDy algorithm\footnote{\revb{For a good overview of recent developments in weak form methods, see \mbox{\cite{MTDB,BMT}.}}}. Section \mbox{\ref{sec:ex_lib}} provides an example of a candidate library while Section \mbox{\ref{sec:strong_v_weak}} shows how sparse regression is accomplished using the weak form. Section \mbox{\ref{sec:hyp_select}} highlights some important hyperparameters and Section \mbox{\ref{sec:sparsity}} provides detail on enforcing sparsity and the loss function. Finally Section \mbox{\ref{sec:gram}} demonstrates how the Gram matrix is used as a similarity measure on the candidate libary.}

\subsubsection{\revb{Example Candidate Library}}\label{sec:ex_lib}

As an example, consider the 1D Euler equation given by Equation \ref{eq:E} by dropping the magnetic field terms. In one dimension, the quantities considered would be energy $E$, velocity $v_x$, density $\rho$, and pressure $p$ (equivalent to total pressure $p^*$ in the absence of a magnetic field). A candidate library consisting of second order monomials and only first order derivatives would contain the terms in Equation \ref{eq:lib_example}. The candidate libraries used in practice for the problems presented here are described in a similar manner in Tables \ref{tab:BW} and \ref{tab:2D}.

\newcommand*{\horzbarl}{\quad\rule[.5ex]{2.5ex}{0.5pt}}
\newcommand*{\horzbarr}{\rule[.5ex]{2.5ex}{0.5pt}\quad}
\begin{equation}
    \Theta = \partial_x
    \left[
    \begin{array}{*{14}c}
        | & | & | & | & | & | & | & | & | & | & | & | & | \\
        E & v_x & \rho & p & E^2 & E v_x & E \rho & Ep & v_x^2 & v_x p & \rho^2 & \rho p & p^2 \\
        | & | & | & | & | & | & | & | & | & | & | & | & |
    \end{array}\label{eq:lib_example}
    \right]
\end{equation}

\subsubsection{\revb{Model Identification in Strong vs Weak PDE Form}}\label{sec:strong_v_weak}

WSINDy uses the same approach as SINDy, finding as few terms as possible from a candidate library that matches the data well to approximate the function $F$ in Equation \ref{eq:strong}, however it performs its analysis using the weak form of the PDE. Given $S$ differential operators $(D^{\alpha^s})_{s \in \lbrace S \rbrace}$, $K$ test functions $(\psi_k(x,t))_{k \in \lbrace K \rbrace}$, and $J$ trial functions $(f_j(x,t))_{j \in \lbrace J \rbrace}$, the PDE in Equation \ref{eq:strong} can be re-written in weak form as a linear system as in Equation \ref{eq:weak}. Comparing the weak form in Equation \ref{eq:weak} to the strong form in Equation \ref{eq:sindy}, the vector $b$ acts in place of $\partial_t u(x,t)$ and matrix $G$ takes the place of the candidate library $\Theta$. WSINDy then solves for sparse $\mathbf{w}^*$, the vector of learned coefficients for terms in the candidate library. Thus the equation form for WSINDy is still a linear matrix system, just built on the weak form of the PDE instead of the strong form. Note $\alpha^s$ represents a multi-index for compactly describing derivatives. The processes for constructing the test functions, sparsely solving for $\mathbf{w}^*$, and more detail on deriving the weak-form linear system are described in full detail in \cite{wsindy_pde}.

\begin{equation}
    \begin{split}
        b &= G \mathbf{w}^* \\
        b_k &= \int\limits_0^T \int\limits_{\Omega} \partial_t (\psi_k(x,t)) u(x,t) \\
        G_{k,(s-1)J+j} &= \int\limits_0^T \int\limits_{\Omega} (-1)^{|\alpha^s|}D^{\alpha^s}\psi_k(x,t) f_j(u(x,t))
    \end{split}
    \label{eq:weak}
\end{equation}

\subsubsection{\revb{Hyperparameter Selection}}\label{sec:hyp_select}
There are a number of hyper-parameters regarding selection and construction of the test functions. While these can be set manually, there are heuristics for automatically selecting some of these values. Summarizing briefly from \cite{wsindy_pde}, the test functions $\psi_k(x,t)$ are piecewise-polynomials that are assumed to be separable in space and time. Test functions are centered about query points $(x_k,t_k)$. The number and locations of these query points are selected by considering the desired query point spacing $s$ and the desired test function support widths $m$. The support lengths $m$ are found such that they ensure the test functions properly decay in both Fourier and real space to ensure compact support (for more details see Appendix A of \cite{wsindy_pde}). The sparsity parameter $\lambda$ is identified by searching over a range of potential values and selecting the one that minimizes the loss function defined in Equation \revb{\mbox{\ref{eq:wsindy_loss}}}. Terms with coefficients and magnitudes (relative to b) less than $\lambda$ are set to zero during the \revb{modified sequential-thresholding least-squares algorithm (MSTLS) process described in Section \mbox{\ref{sec:sparsity}}}. The loss function, given by Equation \revb{\mbox{\ref{eq:wsindy_loss}}}, accounts for what percent of the candidate library has non-zero coefficients (is included in the learned PDE) to balance simplicity with accuracy.

\subsubsection{\revb{Enforcing Sparsity}}\label{sec:sparsity}
\revb{Once the candidate library has been computed as in matrix $G$ in Equation \mbox{\ref{eq:weak}}, the modified sequential-thresholding least-squares algorithm (MSTLS) is followed to enforce sparsity in the resulting PDE. In this process, the least squares solution to $b = G w$ is computed. Given a sparsity setting $\lambda$ upper and lower bounds on learned coefficients $w$ are used to eliminate library terms with insignificant coefficients. These terms are then removed from the library, and the least squares solution of the remaining terms computed. This process continues (solving least squares then removing candidate terms with small coefficients) until no more terms can be eliminated.

This is repeated for a list of sparsity values where the loss at each sparsity setting is calculated as in Equation \mbox{\ref{eq:wsindy_loss}}.}

\begin{equation}
    L(\lambda) = \frac{||G(w^{\lambda} - w^{LS})||_2}{||Gw^{LS}||_2} + \frac{\#I^{\lambda}}{SJ}
    \label{eq:wsindy_loss}
\end{equation}

\revb{Note here $w^{LS}$ is the initial least squares solution to $b=Gw$ while $w^{\lambda}$ is the solutino found by the MSTLS process using sparsity setting $\lambda$. $\#I^{\lambda}$ is the number of nonzero terms in $w^{\lambda}$ and $SJ$ is the total number of terms in the candidate library. The second term in the loss then represents the fraction of terms from the original candidate library that are included in the learned solution. The optimal sparsity setting $\lambda$ is then selected using this loss value. Thus the MSTLS algorithm enforces sparsity in the solution while the loss function in Equation \mbox{\ref{eq:wsindy_loss}} is used to select the optimal sparsity level.}

\subsubsection{\revb{Similarity Analysis using Gram Matrix}}\label{sec:gram}
The matrix $G$ in Equation \ref{eq:weak} contains the candidate library terms $f_j(x,t)$ convolved with the test functions $\psi_k(x,t)$. Calculating the absolute value of $G^TG$, which is referred to as the Gram matrix, provides a description of how numerically similar the trial functions look once in the weak form. Dividing by the maximal value makes all values between zero and one, where one corresponds to a trial function compared to itself, so the lesser the value in this matrix the more unique those candidate terms are when compared to each other. This is equivalent to normalizing the columns of $G$ first. Looking at the off-diagonal values in this matrix then gives a description of the numerical uniqueness of each library term, with one indicating perfect correlation and zero indicating the terms are orthogonal.

\begin{figure}
    \centering
    \includegraphics[width=0.95\textwidth]{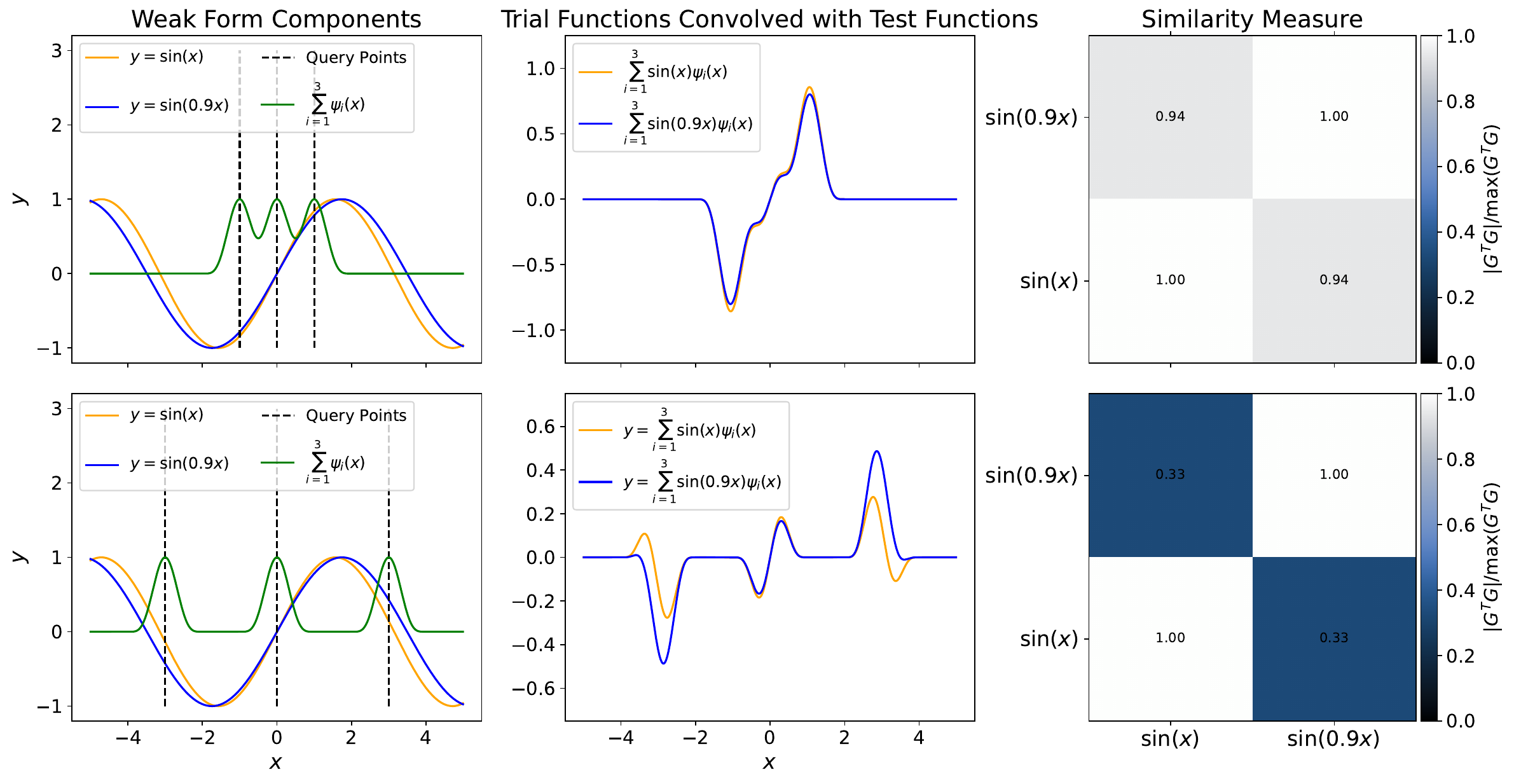}
    \caption{Example showing the significance of $|G^TG|$ as related to the weak construction used in Equation \ref{eq:weak}. The two candidate library terms considered are $\sin(x)$ and $\sin(0.9x)$, shown in the blue and orange curves in the left column. The green curves in the left column represent possible placements of test function $\psi(x)$ centered about the vertical dashed lines denoting the query points. The middle column shows what the two \revc{trial functions} look like once integrated against the test functions. The resulting normalized $|G^TG|$ calculation is shown in the right column. Notice that when the test functions encompass a region of the $x$ domain where the two $\sin$ curves look similar, as in the top row, the resulting off-diagonal quantities in $|G^TG|$ are near one. Whereas in the bottom row, where the test functions cover less similar regions of the $\sin$ curves, the off diagonal terms of $|G^TG|$ are closer to zero.}\label{fig:GTG}
\end{figure}

For example consider a candidate library containing $\sin(x)$ and $\sin(0.9x)$ as shown in the left column of Figure \ref{fig:GTG}. The dashed black lines denote example query points with corresponding test functions in green centered on those points. Looking over a large range of x values it's easy to see that these two functions are different. To construct matrix $G$ in Equation \ref{eq:weak}, the inner product of each \revc{trial function} (here $\sin(x)$ and $\sin(0.9x)$) with the test functions is calculated, resulting in the lines in the center column of Figure \ref{fig:GTG}. The quantities in $G$ are these values integrated over $x$. Calculating $|G^TG|$ and normalizing results in the matrices in the right most column. Notice in the top row, the test functions with compact support only cover a region of the two sine curves where they are close in value. This results in the two sine functions appearing similar in the weak form, as shown in the top figure in the middle column. Their similarity measure of $0.94$ found in $|G^TG|$ is close to one (the value found when the two functions compared are exactly equal). However, in the bottom row of Figure \ref{fig:GTG} the test functions encompass regions where the functions are more obviously different. This results in a similarity measure from $|G^TG|$ of $0.33$. In this way the values in $|G^TG|$ represent how similar \revc{trial functions} are once integrated against the test functions with a value of one meaning the two appear identical and a value of zero meaning they are easily distinguished.

\subsection{Shannon Information Entropy}\label{sec:sie}

When using WSINDy to identify governing PDEs, it would be useful to have a metric that helps indicate whether the method is well-informed by the data regarding the PDEs we are attempting to identify. One way of thinking about how successful WSINDy is at identifying a PDE is whether or not there is enough information in the data for a given PDE to be extracted. That is, we hope Shannon information entropy can quantify how much information is contained in the variable we're trying to find a PDE for.

Shannon information entropy is a measure of the amount of information in a signal needed to identify one disjoint outcome from others \cite{ML_data_2008,engelmann_towards_2022}. For an outcome $X_i$ with probability $P(X_i)$, the amount of information for a single outcome is defined as $I(X_i)$ in Equation \ref{eq:info}. The information entropy for that system $H(X)$ can then be defined as in Equation \ref{eq:ent} \cite{ML_data_2008}.

\begin{subequations}
    \begin{equation}\label{eq:info}
        I(X_i) = -\log (P(X_i))
    \end{equation}
    \begin{equation}\label{eq:ent}
        H(X) = -\sum\limits_{i}^{m} P(X_i)\log(P(X_i))
    \end{equation}
\end{subequations}

For the simulations considered here, $X$ is a quantity evolved by a governing PDE that we wish to recover. For example, consider Equation \ref{eq:cont}. Here $X_i$ represents a single value of density $\rho$. To build the probability distribution $P(X)$, all of the $\rho$ values from a dataset (across all time and spatial locations) are used to build a normalized histogram. The resulting probability distribution function represents how frequently each observed density quantity $\rho$ is encountered in a single simulation.

\begin{figure}[t!]
    \centering
    \includegraphics[width=0.45\textwidth]{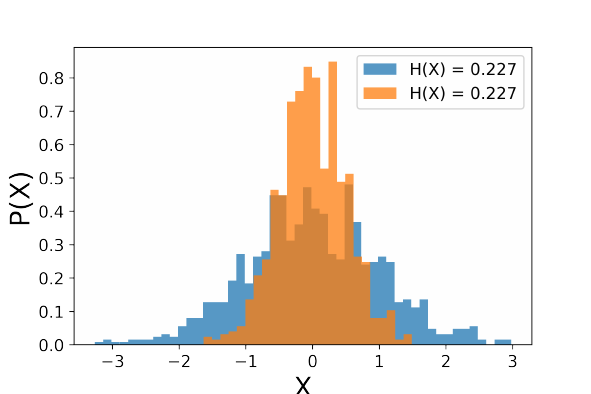}
    \caption{Example of entropy calculation for data sampled from a Gaussian distribution. The orange data is over the original values and the blue data is the same samples just with their values doubled. Thus the two probability distributions are the same shape, just over a different range of values. Regardless of the number of bins used or the magnitude of the values, the information entropy only depends on the shape of the distribution, and is therefore equal between the blue and orange distributions.}\label{fig:H_example}
  \end{figure}

Information entropy only depends on the shape of the distributions and not the magnitude of the values nor widths or number of discrete bins \cite{engelmann_towards_2022}. Consider a simulation with a Gaussian distribution of values. Now consider if all those values are doubled. The two samples contain the same information up to a multiple of a constant, so their entropies should be the same. This is ensured by normalizing the data to mean zero an unit variance before building the PDF, making sure the area under the PDF is one, and dividing the final entropy value by the number of bins. The result, as shown in Figure \ref{fig:H_example}, demonstrates that regardless of the number of bins or re-scaling of values the calculated information entropy remains the same.

In relation to WSINDy, \revb{the original data values are used to calculate the entropy while those same values are integrated against the test functions to build the vector $b$ in Equation \mbox{\ref{eq:weak}}}, which is the vector WSINDy sparsely fits to. The information entropy is used as a measure of the information contained in the data used to construct $b$, and thus reflects the complexity of what is going on in the data provided to WSINDy.

%% file: Exp.tex

A range of ideal magnetohydrodynamics (MHD) test problems are discussed in this paper to explore limitations of WSINDy on different initial conditions in both one and two spatial dimensions. All data discussed in this paper is generated by Athena++, an adaptive mesh refinement code for MHD simulations \cite{athena++}. WSINDy is used to recover the equations for density $\rho$, momentum $\rho \mathbf{v}$, magnetic field $\mathbf{B}$, and energy $E$ in the form of Equations \ref{eq:cont} through \ref{eq:E}. The pressure $p$ is defined as in Equation \ref{eq:p_def} while the total pressure $p^*$ takes the form of Equation \ref{eq:ptot}.

\begin{subequations}
  \begin{equation}\label{eq:cont}
      \partial_t\rho + \nabla \cdot (\rho \mathbf{v}) = 0
  \end{equation}
  \begin{equation}\label{eq:mom}
      \partial_t(\rho\mathbf{v}) + \nabla \cdot [\rho\mathbf{v}\mathbf{v} - \mathbf{B}\mathbf{B} + p^*] = 0
  \end{equation}
  \begin{equation}
      \partial_t\mathbf{B} - \nabla \times (\mathbf{v} \times \mathbf{B}) = 0
  \end{equation}
  \begin{equation}\label{eq:E}
      \partial_tE + \nabla \cdot \bigg[ (E + p^*)\mathbf{v} - \mathbf{B}(\mathbf{B} \cdot \mathbf{v}) \bigg] = 0
  \end{equation}
  \begin{equation}\label{eq:p_def}
    p = (\gamma - 1)(E - \frac{1}{2}\rho(\mathbf{v} \cdot \mathbf{v}) - \frac{1}{2}(\mathbf{B} \cdot \mathbf{B}))
 \end{equation}
  \begin{equation}\label{eq:ptot}
     p^* = p + \frac{1}{2}(\mathbf{B} \cdot \mathbf{B})
  \end{equation}
  \label{eq:ideal_mhd}
\end{subequations}


We consider four problems commonly used for testing MHD codes. Section \ref{sec:riemann} introduces two Riemann problems (the Brio-Wu Shock Tube and the Magnetized Sedov Blast) whose initial conditions consist of discontinuities. The remaining datasets are 2D simulations with more complex initial conditions. Section \ref{sec:OT} discusses the Orszag-Tang Vortex which contains increasingly small scale interactions. Finally Section \ref{sec:cloud} introduces the Shock Cloud problem, where a velocity front collides with an over-dense cloud. Each of these problem setups is used to test for different features of simulation codes and therefore each highlight different behaviors in the resulting data. \revb{Details of the simulation settings used with Athena++ can be found in Table \mbox{\ref{tab:athena_settings}}.}

\begin{table}
  \centering
  \begingroup\setlength{\fboxsep}{0pt}
  \colorbox{cyan!\transcolor}{%
  \begin{tabular}{|c|c|c|c|c|c|}
      \hline
      Simulation & Grid Size & Integrator & CFL & $\gamma$ & BCs \\ 
      \hline
      Brio-Wu & 256 & vl2 & 0.4 & 2 & Outflow \\ 
      \hline
      Magnetized Sedov Blast & 128 $\times$ 128 & vl2 & 0.3 & $5/3$ & Periodic \\
      \hline
      Orszag-Tang & 500 $\times$ 500 & vl2 & 0.4 & $5/3$ & Periodic \\
      \hline 
      Shock Cloud & 200 $\times$ 90 & vl2 & 0.3 & $5/3$ & Outflow \\
      \hline
  \end{tabular}}\endgroup
  \caption{\revb{Key simulation parameters for the Athena++ simulations discussed in Section \mbox{\ref{sec:exp}}. The grid size, time integrator, Courant-Friedrichs-Lewy (CFL) number, specific heat ratio $\gamma$, boundary conditions (BCs), and time parameters are described. The Orszag-Tang datasets are run until 10,000 code cycles are reached while the Magnetized Sedov Blast simulations are stopped before the shock front passes the center of the domain, as shown in Figure \mbox{\ref{fig:sedov_fc}}. The Shock Cloud datasets are similarly run until just before the over-dense bubble has left the domain. All simulations use the Athena++ default second-order accurate van Leer integrator with Harten-Lax-van Leer discontinuities (HLLD) riemann solver.}}\label{tab:athena_settings}
\end{table}

\subsection{Riemann Problems}\label{sec:riemann}

Many test problems contain shocks, necessitating model identification codes such as WSINDy that effectively deal with discontinuous data. Riemann problems (RP) are typically employed to test the ability of forward simulation codes to accurately track discontinuities and rarefactions expected from systems of conservation laws \cite{orig_bw,orig_sedov}. A RP consists of a system of conservation laws and piecewise-constant initial conditions \cite{LeVeque}. For example, in one spatial dimension this can be represented as 

\noindent\begin{subequations}
  \noindent\centering
  \begin{minipage}{0.48\textwidth}
      \begin{align}
        \partial_t\mathbf{u} + \partial_x \mathbf{F}(\mathbf{u}) = 0\label{eq:cons_law}
      \end{align}
  \end{minipage}
  \hfill
  \begin{minipage}{0.48\textwidth}
      \begin{align}
        \mathbf{u}_0(x)  = \begin{cases} \mathbf{u}_L, & x<0 \\ \mathbf{u}_R, &x\geq 0.\end{cases} \label{eq:riemann_ic}
      \end{align}
  \end{minipage}\bigskip
\end{subequations}


Consider a system of $n$ conservation laws with flux $\mathbf{F}(\mathbf{u}) = \mathbf{A}\mathbf{u}$ where $\mathbf{A}$ is an $n\times n$ matrix with eigen-decomposition $\mathbf{A} = \mathbf{X}\Lambda \mathbf{X}^{-1}$, and all real eigenvalues $\lambda_i = \Lambda_{ii}$ (i.e.\ the system is hyperbolic). The weak entropy solution to this system consists of $n$ shocks traveling at speeds $\lambda_i$, each with solution jumps $\mathbf{w}_i$ given by multiples of the eigenvectors $\mathbf{X}$. $\mathbf{u}$ is also a solution to any other conservation law with flux $\mathbf{f}$ so long as the Rankine-Hugoniot conditions are still satisfied such that $[\mathbf{f}(\mathbf{u})] = \lambda_i\mathbf{w}_i$ for each of the $i=1,\dots,n$ shocks \cite{LeVeque}. For example, we can let the  $j$th component of $\mathbf{f}$ take the form of a quadratic

\begin{equation}
  \mathbf{f}_j(\mathbf{u}) = \mathbf{u}^T\mathbf{Q}^{(j)}\mathbf{u}
\end{equation}

for matrices $\mathbf{Q}^{(j)}$ given by $\mathbf{Q}^{(j)}_{ik} = \sum_{\ell=1}^n\mathbf{X}_{j\ell}\tilde{\mathbf{A}}_{\ell i}\mathbf{X}_{\ell k}^{-1}$, so long as the rows of $\tilde{\mathbf{A}}$ (denoted by $\tilde{\mathbf{A}}_i$) satisfy $\tilde{\mathbf{A}}_i^T[\mathbf{v}\mathbf{v}_i] = \lambda_i[\mathbf{v}_i]$ for each $i$, where $\mathbf{v} = \mathbf{X}^{-1}\mathbf{u}$ are the decoupled solution components. These polynomial combinations of the various solutions $\mathbf{u}$ to these conservation laws will contain many redundant models, and opportunities for redundancy grow in the number of state variables.



\subsubsection{Brio-Wu Shock Tube}\label{sec:BW}
\begin{figure}[t!]
  \centering
  \begin{subfigure}[t]{0.49\textwidth}
    \centering
    \includegraphics[width=0.9\textwidth]{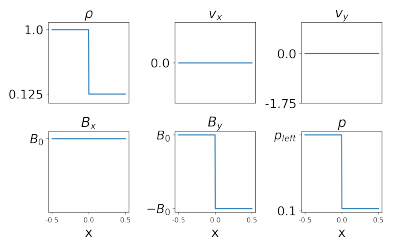}
    \caption{Example of initial conditions for Brio-Wu Shock Tube with varied parameters $p_{left}$ and $B_0$. The $x$-axis ranges from -0.5 to 0.5 with discontinuities in magnetic field, density, and pressure occurring at $x=0$.}\label{fig:bw_ic}
  \end{subfigure}%
  ~ 
  \begin{subfigure}[t]{0.49\textwidth}
    \centering
    \includegraphics[width=0.9\textwidth]{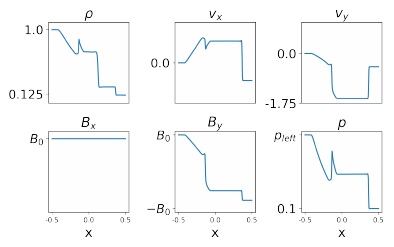}
    \caption{Example of final state for Brio-Wu Shock tube using the initial conditions shown in panel (a). The simulations are stopped before any of the rarefaction waves/fans or shocks escape the outflow boundary conditions \cite{BW_waves}. Note for this problem the x component of magnetic field $B_x$ is constant.}\label{fig:bw_fc}
  \end{subfigure}
  \caption{Figures \ref{fig:bw_ic} and \ref{fig:bw_fc} respectively show the initial and final states of an example Brio-Wu shock tube dataset. The quantities shown are density ($\rho$), x and y velocity components ($v_x$,$v_y$), x and y magnetic field components ($B_x$,$B_y$), and pressure ($p$). The varied parameters considered are $p_{left}$, representing the size of the pressure jump, and $B_0$, the magnetic field strength.}\label{fig:bw_data}
\end{figure}

The Brio-Wu Shock Tube is a one and a half dimensional problem originally used to test the robustness of a Roe-type upwind difference scheme to shocks for the ideal MHD equations \cite{orig_bw}. The initial features of the dataset are discontinuities at $x=0$ in density $\rho$, y-magnetic field $B_y$, and pressure $p$, while the remaining quantities are constant as shown in Figure \ref{fig:bw_ic}. Variations of the Brio-Wu Shock Tube are considered here wherein the size of the jumps in pressure and magnetic field are changed \cite{athena++,orig_bw}. These variations are discussed in terms of $B_0$, representing the magnitude of the magnetic field, and $p_{left}$, representing the kinetic pressure on the left side of the jump. The initial pressure to the right of the discontinuity is always set to $0.1$ while $p_{left} \geq 0.1$. This parameterization relative to the initial and final states is shown in Figure \ref{fig:bw_data}. A total of 399 variations of these initial conditions are considered by varying $p_{left}$ from $0.1$ (no pressure jump in initial conditions) to $1.0$ in steps of $0.05$ and similarly $B_0$ ranges from $0$ to $1.0$ in steps of $0.05$. 

\subsubsection{Magnetized Sedov-Taylor Blast}\label{sec:sedov}
\begin{figure}[t!]
  \centering
  \begin{subfigure}[t]{0.49\textwidth}
    \centering
    \includegraphics[width=0.9\textwidth]{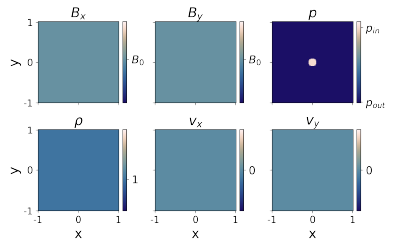}
    \caption{Initial conditions for Magnetized Sedov-Taylor Blast with varied parameters denoted by $B_0$ for the magnetic field and $p_{in}$ for the increased kinetic pressure of the bubble \cite{orig_sedov, sedov_desc}. $p_{out}$ is always set to $0.1$.}\label{fig:sedov_ic}
  \end{subfigure}%
  ~ 
  \begin{subfigure}[t]{0.49\textwidth}
    \centering
    \includegraphics[width=0.9\textwidth]{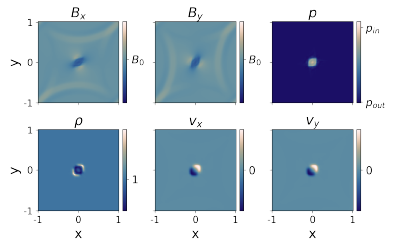}
    \caption{Example of final state for Magnetized Sedov Blast using the initial conditions shown in panel (a). The simulations are stopped so the shock front does not reach the center of the simulation domain after passing through the periodic boundaries.}\label{fig:sedov_fc}
  \end{subfigure}
  \caption{Figures \ref{fig:sedov_ic} and \ref{fig:sedov_fc} respectively show the initial and final states of an example Magnetized Sedov Blast dataset. The color bars are consistent between the initial and final states. The quantities shown are density ($\rho$), x and y velocity components ($v_x$,$v_y$), x and y magnetic field components ($B_x$,$B_y$), and pressure ($p$). The varied parameters considered are $p_{in}$, representing the pressure inside the bubble, and $B_0$, the magnetic field strength.}\label{fig:sedov_data}
\end{figure}

The Magnetized Sedov-Taylor blast is a two dimensional ideal MHD problem that tests whether a simulation method is divergence-free in the presence of discontinuities, which is often of concern in codes focusing on astrophysical simulations \cite{orig_sedov,sedov_desc}. All quantities are initially constant with the exception of pressure, where a circle of high pressure is placed at the center of the domain as shown in Figure \ref{fig:sedov_ic}. Variations of the Magnetized Sedov-Taylor Blast are considered here wherein the size of the jump in pressure and magnitude of the magnetic field are changed \cite{athena++,orig_sedov}. These variations are discussed in terms of $B_0$, representing the magnitude of the magnetic field, and $p_{in}$, representing the kinetic pressure inside the initial bubble. The notation $p_{ratio}$ will also sometimes be used to denote the ratio of the pressure inside the bubble $p_{in}$ to the ambient pressure outside the bubble $p_{out} = 0.1$. This parameterization is shown in Figure \ref{fig:sedov_ic}. 

\subsection{Smooth Initial Conditions: Orszag-Tang Vortex}\label{sec:OT}
\begin{figure}[t!]
  \centering
  \begin{subfigure}[t]{0.49\textwidth}
    \centering
    \includegraphics[width=0.9\textwidth]{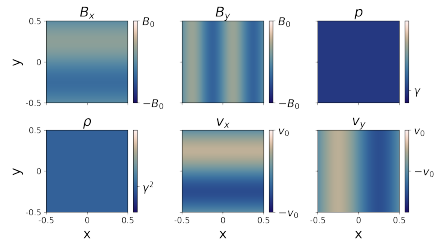}
    \caption{Initial conditions for Orszag-Tang dataset with varied parameters denoted by $B_0$ for the magnetic field and $v_0$ for the magnitude of the initial velocity.}\label{fig:ot_ic}
  \end{subfigure}%
  ~ 
  \begin{subfigure}[t]{0.49\textwidth}
    \centering
    \includegraphics[width=0.9\textwidth]{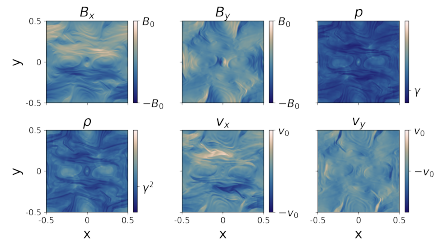}
    \caption{Example of final state for Orszag-Tang using the initial conditions shown in panel (a). The boundary conditions are periodic by design so there is no need for a maximum simulation time to limit escape from the simulation.}\label{fig:ot_fc}
  \end{subfigure}
  \caption{Figures \ref{fig:ot_ic} and \ref{fig:ot_fc} respectively show the initial and final states of an example Orszag-Tang dataset. The simulated quantities are density ($\rho$), x and y velocity components ($v_x$,$v_y$), x and y magnetic field components ($B_x$,$B_y$), and pressure ($p$). The varied parameters considered are $v_0$, representing the initial maximal velocity magnitude, and $B_0$, the magnetic field strength.}\label{fig:ot_data}
\end{figure}

The Orszag-Tang problem is commonly used to test a code's robustness to shock-shock interactions, whether the divergence of the magnetic field is zero, and whether symmetry is preserved \cite{OT}. These initial conditions are varied through the coefficients $B_0$ and $v_0$ associated with the initial states of the magnetic field and velocity respectively, as defined in Equations \ref{eq:OT_1} through \ref{eq:OT_2} and shown in Figure \ref{fig:ot_data}. The parameters $v_0$ and $B_0$ range from $0$ to $1.0$ in steps of $0.05$, amounting to 440 datasets. \revb{Data is subsampled evenly in space so WSINDy only uses every 5\textsuperscript{th} data point in each spatial dimension, amounting to 4\% of the overall data.}

\noindent\begin{subequations}
  \noindent\centering
  \begin{minipage}{0.48\textwidth}
      \begin{align}
        \rho(x,y,t=0) &= \gamma^2 \label{eq:OT_1} \\
        v_x(x,y,t=0) &= -v_0\sin(y) \\
        B_x(x,y,t=0) &= -B_0\sin(y)
      \end{align}
  \end{minipage}
  \hfill
  \begin{minipage}{0.48\textwidth}
      \begin{align}
        p(x,y,t=0) &= \gamma \\
        v_y(x,y,t=0) &= v_0\sin(x) \\
        B_y(x,y,t=0) &= B_0\sin(2x) \label{eq:OT_2} 
      \end{align}
  \end{minipage}\bigskip
\end{subequations}


\subsection{Interacting Discontinuities: Magnetized Shock Cloud Collision}\label{sec:cloud}
\begin{figure}[t!]
  \centering
  \begin{subfigure}[t]{0.49\textwidth}
    \centering
    \includegraphics[width=0.985\textwidth]{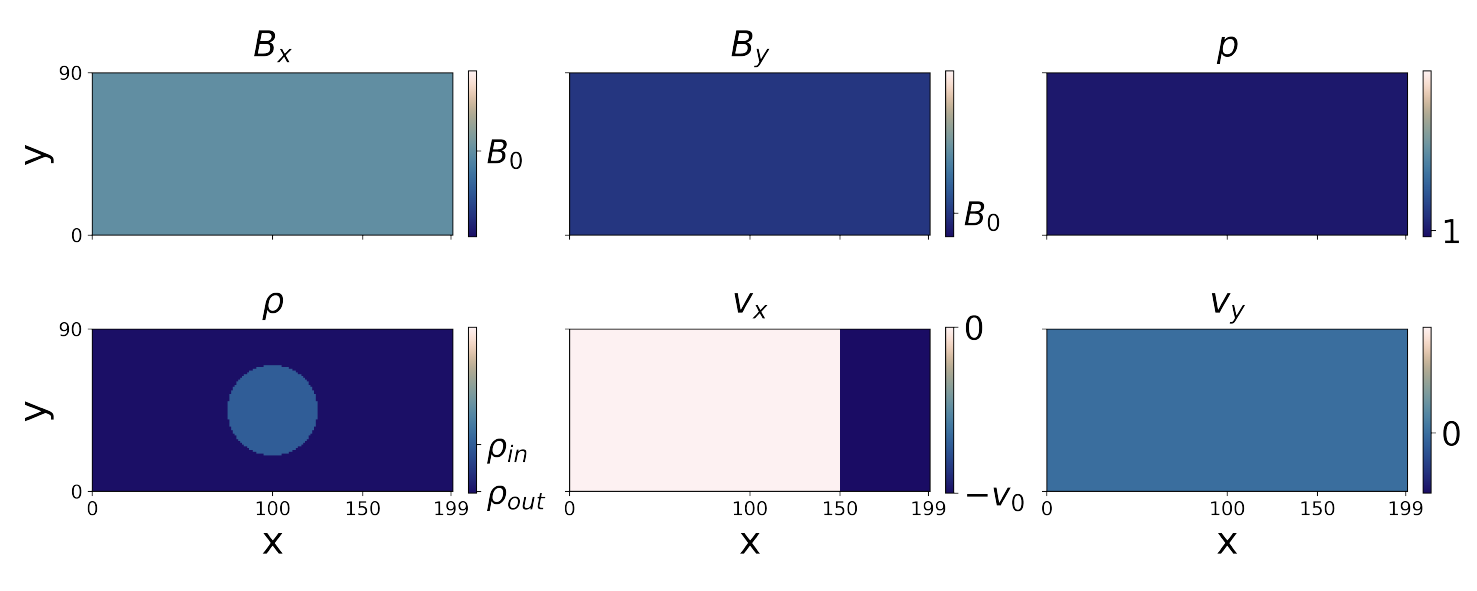}
    \caption{Initial conditions for magnetized shock bubble with varied parameters denoted by $B_0$ for the magnetic field and $v_0$ for the magnitude of the velocity front. The magnetic field is constant and split equally between components $B_x$ and $B_y$, meaning the total magnetic field is at a 45 degree angle.}
    \label{fig:sb_ic}
  \end{subfigure}%
  ~ 
  \begin{subfigure}[t]{0.49\textwidth}
    \centering
    \includegraphics[width=0.985\textwidth]{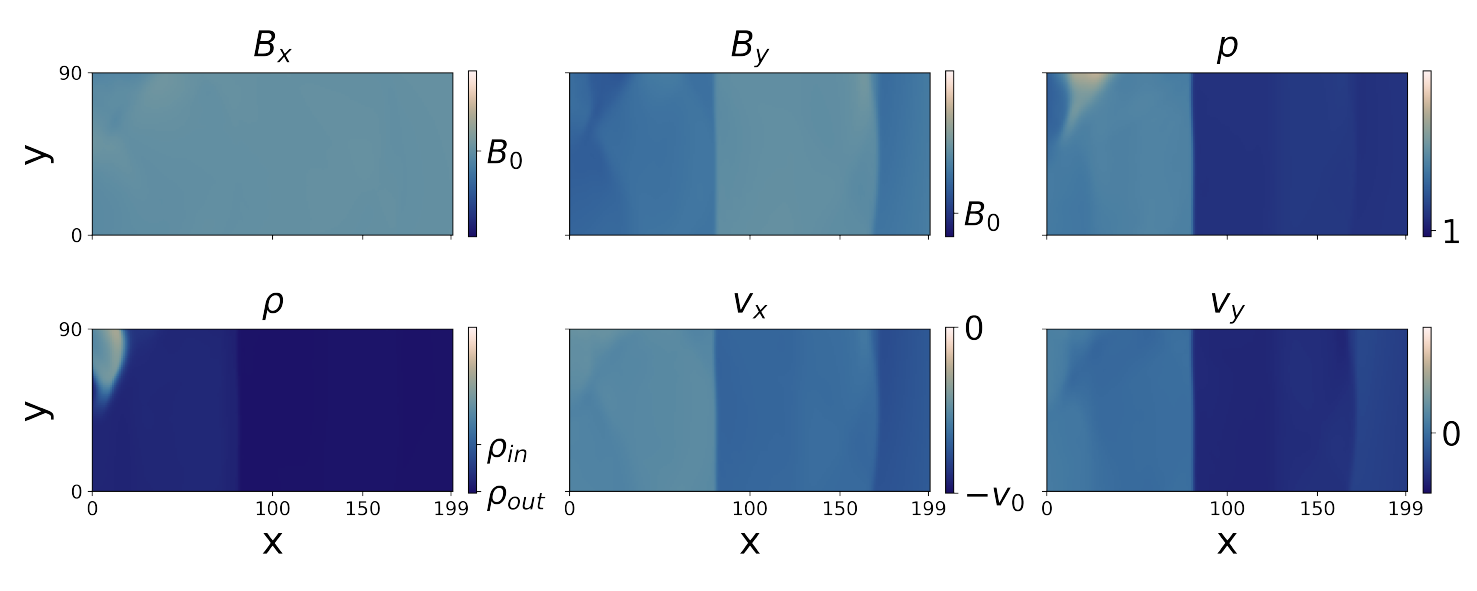}
    \caption{Example of final state for magnetized shock cloud using the initial conditions shown in panel (a). The simulations are stopped before the dense bubble is fully pushed through the outflow boundary conditions by the velocity front.}\label{fig:sb_fc}
  \end{subfigure}
  \caption{Figures \ref{fig:sb_ic} and \ref{fig:sb_fc} respectively show the initial and final conditions of an example magnetized shock bubble dataset. The quantities shown are density ($\rho$), x and y velocity components ($v_x$,$v_y$), x and y magnetic field components ($B_x$,$B_y$), and pressure ($p$). The varied parameters considered are $\rho_{in}$, representing the initial cloud density, and $B_0$, the magnetic field strength.}\label{fig:sb_data}
\end{figure}

The Magnetized Shock Cloud Collision is a variation of a set of initial conditions used in astrophysics to study how a shock wave interacts with a cloud of a different density than the ambient medium, resulting in a complex set of shock and wave interactions 
\cite{mhd_shock_cloud,shock_interstellar_cloud,cassiopeia}. The entire domain is initially at constant pressure with a circular cloud of density $\rho_{in}$ and a shock of velocity $v_0$ as shown in Figure \ref{fig:sb_ic}. The initial magnetic field is at a 45 degree angle in the simulation domain so the x and y magnetic field components, $B_x$ and $B_y$, are initially both equal to $B_0$, as shown in Figure \ref{fig:sb_ic}. This results in the dense cloud moving toward the top left corner of the domain, as shown in the lower left tile of Figure \ref{fig:sb_fc}. The cloud density is set at $\rho_{in} = 5$ compared to the ambient density $\rho_{out} = 1$ with $v_0$ and $B_0$ ranging from $0$ to $10$ in steps of $1$, amounting to 121 datasets. The maximum simulation time is $200/(0.75v_0)$ so the region of dynamics (the shock interacting with the over dense cloud) does not fully escape the domain.

\subsection{WSINDy Hyperparameters}
WSINDy has a number of hyperparameters that must be set manually or selected algorithmically. For the problems presented here, each equation is recovered independently. The query points spacing $s$ is always set to $5$ \revc{so the points are not so spread apart that dynamics are omitted. Incorporating optimal query points selection into WSINDy and other similar algorithms is an active area of research \mbox{\cite{wsindy_pde,wsindy_ode}}.} The candidate library consists of monomial combinations of state variables up to the expected maximum order for the corresponding ideal MHD equation (as written in Equations \ref{eq:cont} through \ref{eq:E}). Some terms are excluded, as described in Tables \ref{tab:BW} and \ref{tab:2D}, to avoid having two theoretically equivalent forms accounted for in the candidate library. For example, pressure and thermal energy density can be re-written in terms of each other so these terms are only simultaneously included in the candidate library when the equation form calls for both. Only zeroth and first order derivatives are considered. The sparsity parameter $\lambda$ is set by searching over potential values from $10^{-6}$ to $1$.

\begin{table}
  \centering
  \begin{tabular}{|c|c|c|}
      \hline
      Equation & Max Monomial Order & Excluded Trial Functions \\
      \hline
      $\partial_t(\rho)$ & 2 & $p^*$, $E$, $B_x$ \\
      \hline
      $\partial_t(\rho v_x)$ & 3 & $(p^*)^z$ s.t. $z>1$, $E$, $B_x$ \\
      \hline
      $\partial_t(\rho v_y)$ & 3 & $p^*$, $E$, $B_x$ \\
      \hline
      $\partial_t(B_y)$ & 2 & $p^*$, $E$, $B_x$\\
      \hline
      $\partial_t(E)$ & 3 & $(p^*)^z$ s.t. $z>1$, $(E)^z$ s.t. $z>1$, $B_x$\\
      \hline
  \end{tabular}
  \caption{Candidate trial functions, corresponding to $f_j(x,t)$ in Equation \ref{eq:weak}, considered when running WSINDy on 1.5D Brio-Wu datasets. Terms containing $B_x$ are always removed as in this problem setup as it is constant and therefore not a dynamic variable. The excluded trial functions column specifies that trial functions containing the listed terms are excluded from the candidate library. For example, for the continuity equation $\partial_t(\rho)$ all terms containing total pressure $p^*$ are excluded. For the x-momentum equation $\partial_x(\rho v_x)$, all terms containing total pressure $p^*$ to a power greater than one are excluded. The variable $z$ in the excluded terms column represents a power.}\label{tab:BW}
\end{table}

\begin{table}
  \centering
  \begin{tabular}{|c|c|c|}
      \hline
      Equation(s) & Max Monomial Order & Excluded Trial Functions \\
      \hline
      $\partial_t(\rho)$ & 2 & $p^*$, $E$\\
      \hline
      $\partial_t(\rho v_x)$, $\partial_t(\rho v_y)$ & 3 & $(p^*)^x$ s.t. $x>1$, $E$\\
      \hline
      $\partial_t(B_x)$,$\partial_t(B_y)$ & 2 & $p^*$, $E$\\
      \hline
      $\partial_t(E)$ & 3 & $(p^*)^x$ s.t. $x>1$, $(E)^x$ s.t. $x>1$\\
      \hline
  \end{tabular}
  \caption{Candidate trial functions, corresponding to $f_j(x,t)$ in Equation \ref{eq:weak}, considered when running on 2D Ideal-MHD datasets presented in Sections \ref{sec:OT} through \ref{sec:cloud}. Trial functions that contain the terms specified in the excluded trial functions column are excluded from the candidate library. The variable $x$ in the excluded terms column represents a power.}\label{tab:2D}
\end{table}

%% file: Results.tex
Each of the datasets introduced in the simulation data section are analyzed here using a combination of equation recovery by WSINDy, Shannon information entropy, and correlations in the candidate library via the matrix $G$. These three pieces help illuminate what behavior WSINDy performs best on. \revc{Noise is added to each dataset to create slight variations between learned equations and analyze trends. The more information rich the dataset, the more noise can be added.} The performance of WSINDy is related to the information present in the data as well as the uniqueness and therefore distinguishability of terms in the candidate library.

\subsection{Brio-Wu Shock Tube}
\input{BW_results}

\subsection{Magnetized Sedov-Taylor Blast}
\input{Sedov_results}

\subsection{Orszag-Tang Vortex}
\input{OT_results}

\subsection{Magnetized Shock Cloud Interaction}
\input{SC_results}

%% file: BW_results.tex
\begin{figure}[t!]
    \centering
    \begin{subfigure}[t]{0.59\textwidth}
        \centering
        \includegraphics[width=\textwidth]{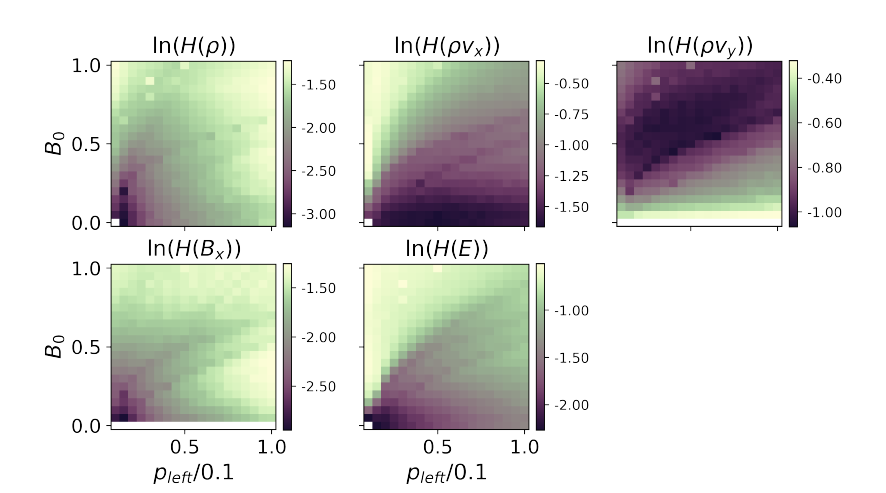}
        \caption{Information entropy calculations for density $\rho$, momentum $\rho v_x$ and $\rho v_y$, magnetic field $B_y$, and energy $E$ of the Brio-Wu datasets. The colorbars differ between plots as the Shannon information entropy is on disparate scales when compared between them.}
        \label{fig:BW_ent}
    \end{subfigure}%
    ~ 
    \begin{subfigure}[t]{0.4\textwidth}
        \centering
        \includegraphics[width=\textwidth]{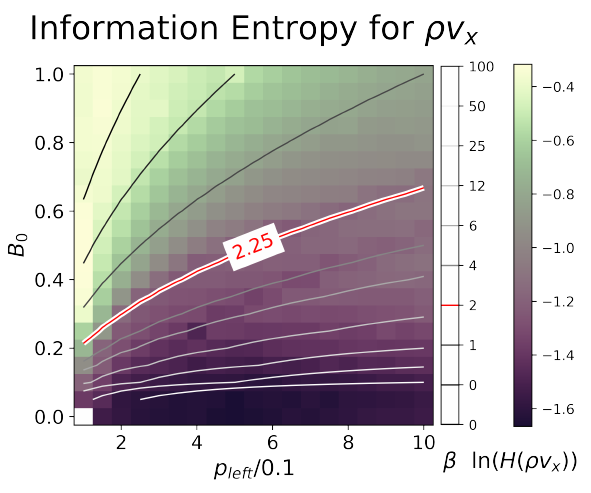}
        \caption{Contours of plasma beta $\beta = p_{left}/B_0^2$ imposed on plot of information entropy $\ln(H(\rho v_x))$ from Figure \ref{fig:BW_ent}. The red contour at $\beta \approx 2.25$ aligns with a division in more and less succesful equation recovery shown in Figure \ref{fig:bw_recoveries} and a dip in entropy observed for all quantities in Figure \ref{fig:BW_ent}.}
        \label{fig:bw_ent_momx}
    \end{subfigure}
    \caption{Shannon information entropy calculated on the Brio-Wu datasets described in Section \ref{sec:BW}. Corresponding to the initial conditions shown in Figure \ref{fig:bw_data}, the x-axis describes an increasing pressure jump $p_{left}$ in the initial conditions and the y-axis represents increasing magnetic field strength $B_0$. The trends in entropy follow the contours of plasma $\beta$. Figure \ref{fig:bw_ent_momx} highlights this alignment using x-momentum $\rho v_x$ from the top center panel of Figure \ref{fig:BW_ent} as an example.}\label{fig:bw_ent_both}
\end{figure}

Figure \ref{fig:BW_ent} shows the computed entropy for each ideal MHD equation over the described variations of the Brio-Wu initial conditions. Generally the entropy increases as the magnetic field strength and pressure jump are increased, however there are trends that correspond to the approximate plasma $\beta$ of the system. Overlaying contours of plasma $\beta$, approximated as $\beta \approx p_{left}/B_0^2$, on the x momentum entropy plot, as shown in Figure \ref{fig:bw_ent_momx}, highlights that the entropy is constant along the plasma beta contours. Furthermore, all quantities show a dip in entropy along the line $\beta = 2.25$, highlighted in red in Figure \ref{fig:bw_ent_momx}. Alternatively the entropy could be calculated on the vector $b$ from Equation \ref{eq:weak} as this is what WSINDy actually fits to rather than the original data itself. Repeating the same calculation on the vector $b$ does not reveal any trends, suggesting this dip in entropy is linked to properties of the original data and is not brought about by the integration with test functions from the weak formulation.

\begin{figure*}
    \centering
    \includegraphics[width=0.65\textwidth]{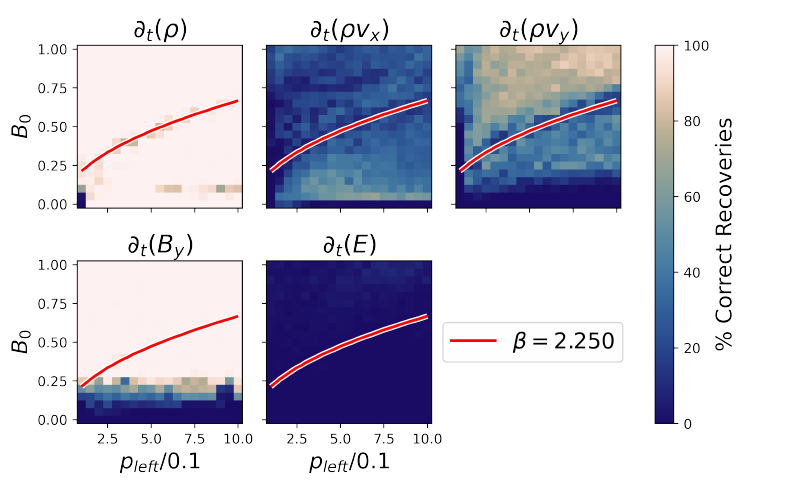}
    \caption{Ideal MHD equation recoveries for 200 repetitions of WSINDy per Brio-Wu data set with 1\% random noise added. The axes denote the values used in the parameterized initial conditions described in Section \ref{sec:BW}. The x-axis is the ratio of the pressure left of the discontinuity $p_{left}$ to the pressure right of the discontinuity $0.1$ in the initial state. The y-axis is the magnetic field strength $B_0$ in the initial state. The red line denotes the plasma beta contour corresponding to that in Figure \ref{fig:bw_ent_momx}. This line of plasma $\beta$ divides regions of good and bad recovery rates for the momentum equations. There is also a dip in recovery rate for the continuity equation along this line. The energy equation is only recovered a handful of times for strong magnetic field $B_0$.}
    \label{fig:bw_recoveries}
\end{figure*}

Using the same datasets as in Figure \ref{fig:BW_ent}, repeated PDE identifications using WSINDy were completed for each dataset. A 1\% noise ratio was used to provide some variation between repeated trials. The results of 200 repeated PDE identifications using WSINDy per dataset are shown in Figure \ref{fig:bw_recoveries}. This is an unnecessarily large number of WSINDy repetitions to examine the recovery trends, but for a problem with one dimension is computationally inexpensive to complete. Correct PDE identifications are counted as WSINDy returning all the terms of the expected ideal MHD equations as written in Equations \ref{eq:cont} through \ref{eq:E} without any missing or extra terms. The identified line of plasma $\beta$ from the entropy analysis is carried over to these plots for comparison.

The trends in recovery seen in Figure \ref{fig:bw_recoveries} echo those of the entropy calculations in Figure \ref{fig:BW_ent} with aligning trends for each equation. For example, in regions of lower entropy for the momentum equations WSINDy is more successful in its equation recovery. The division between regions of improved and worsened recovery align with the line of $\beta = 2.25$ and the dip in entropy for each plot in Figure \ref{fig:BW_ent}. This same line of plasma beta follows a dip in recovery for the continuity equation. The magnetic field equations are easily recovered once enough magnetic field is present and there are only a few times the energy equation is recovered at higher $B_0$ values.

\begin{figure*}
    \centering
    \includegraphics[width=0.65\linewidth]{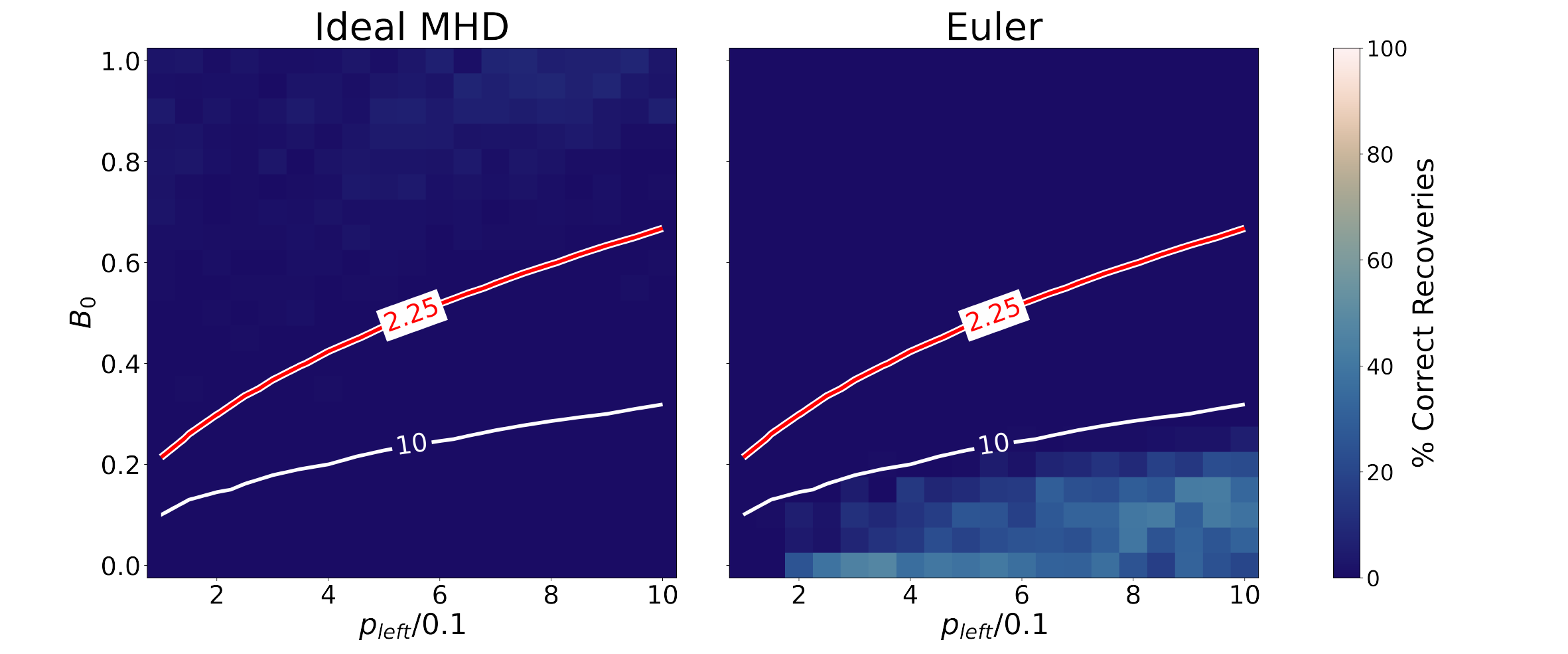}
    \caption{Searches through WSINDy results from Brio-Wu datasets for Ideal-MHD and Euler equation recoveries plus lines of plasma beta. The two plots consider the same set of PDEs identified by WSINDy, but with different criteria for what is considered the expected/correct PDE. At left is the full ideal MHD energy equation, copied from the bottom right panel of Figure \ref{fig:bw_recoveries}, while the right is the energy equation without the magnetic field terms, the Euler equation. The percent of correct recoveries is then out of the repeated calculations on each dataset counting when only all of that equations terms are recovered. The Ideal MHD equation is recovered a few times for high $B_0$ values (strong magnetic field) while the Euler equation is often recovered for datasets with a plasma $\beta$ greater than 10, which is denoted by the white contour.}
    \label{fig:BW_comp_E}
\end{figure*}

Note, however, that the results shown in Figure \ref{fig:bw_recoveries} only consider whether the identified PDEs match the expected ideal MHD form. There are regions of the recovery space that tend to identify a simplified form of the ideal MHD equations, namely the Euler equations. Figure \ref{fig:BW_comp_E} shows the energy equation recoveries from Figure \ref{fig:bw_recoveries} on the left and an altered search for the Euler equations on the right. The change between the two is just what is counted as the correct PDE over the same returned PDEs identified by WSINDy. For initial conditions with a plasma $\beta$ greater than about $10$ (below the white line in Figure \ref{fig:BW_comp_E}) the simplified Euler energy equation tends to be recovered. In this region the influence of the magnetic field is small compared to the other effects so the sparsity aspect of WSINDy returns the ideal MHD energy equation without the magnetic field terms, which is the Euler energy equation. WSINDy identifies the dominant behavior of the system as to return a sparse solution (a PDE with the least terms that still matches the data well enough).

\begin{figure}[t!]
    \centering
    \begin{subfigure}[t]{0.5\textwidth}
        \centering
        \includegraphics[width=\linewidth]{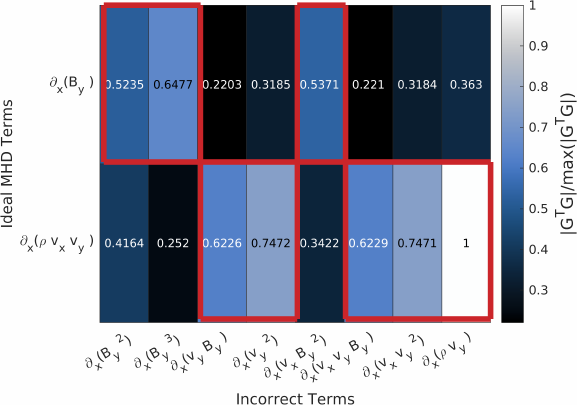}
        \caption{Example of correlation from $|G^TG|$ of most frequently identified unexpected terms to expected ideal MHD terms. The terms along the x-axis are chosen by identifying the most frequently recovered incorrect terms in this region from the same data displayed in Figure \ref{fig:bw_recoveries}. The y-axis terms are those expected from Equation \ref{eq:mom}. Each incorrect term on the x-axis has one expected term to which it is highly correlated. These values are outlined in red and correspond to the vertical dashed lines in Figure \ref{fig:bw_corr_hist}.}\label{fig:bw_corr_mat}
    \end{subfigure}%
    ~ 
    \begin{subfigure}[t]{0.5\textwidth}
        \centering
        \includegraphics[width=\linewidth]{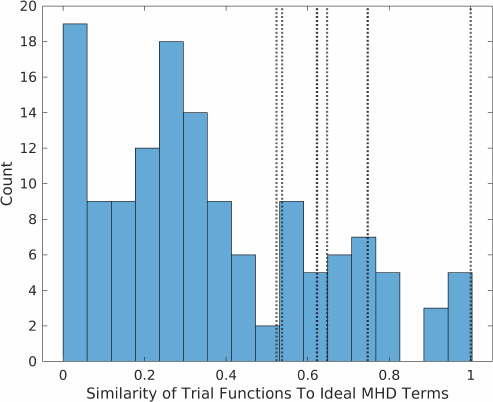}
        \caption{Histogram of similarity of all candidate library terms when compared to the expected terms for this equation. The dashed lines mark the values boxed in red in figure \ref{fig:bw_corr_mat}, denoting the maximum similarity measure of each of the commonly recovered incorrect terms to the expected ideal MHD terms. All the incorrect terms that are encountered reside in the tail of the histogram.}\label{fig:bw_corr_hist}
    \end{subfigure}
    \caption{Analysis of similarity between candidate library terms compared to the expected ideal MHD terms for $\partial_t(\rho v_y)$. These are examples from looking at the Brio-Wu dataset with $p_{left} = 0.7$ and $B_0 = 0.1$, falling in the region of poor recovery for this equation in Figure \ref{fig:bw_recoveries}. All values shown here are calculated from $|G^TG|$ as introduced in Section \ref{sec:sie}.}\label{fig:bw_corr}
\end{figure}

By counting how frequently each term is identified by WSINDy from the repeated runs shown in Figure \ref{fig:bw_recoveries}, different terms are most frequently recovered in different regions of parameter space. The similarity of these unexpected terms to the known ideal MHD terms can then be observed by calculating $|G^TG|$. Recall this is a representation of similarity between candidate library terms once convolved with the test functions. For example, for the y-momentum equation, the expected ideal MHD equation is consistently recovered above the red line of plasma $\beta$ in Figure \ref{fig:bw_recoveries}, but less consistently below the line. Taking one example dataset below the red line of plasma $\beta$ ($p_{left} = 0.7$ and $B_0 = 0.1$), we identify the most commonly recovered terms outside the ideal MHD equations. The most frequently recovered but incorrect terms are shown along the x-axis in Figure \ref{fig:bw_corr_mat}. The theoretically known ideal MHD terms are shown on the vertical axis. The matrix values are found by calculating $|G^TG|$, meaning terms with a value closer to one look numerically similar in the weak representation used by WSINDy and it is therefore difficult to distinguish between them. All of the frequently recovered incorrect terms have a correlation to an expected term of at least $0.5$. Alternatively, considering all candidate library terms correlation to the expected terms in Figure \ref{fig:bw_corr_hist} shows that the bulk of terms in the candidate library look dissimilar to the expected terms, and thus have correlation values closer to zero. The values boxed in red denote which of the expected terms each unexpected term is most similar too. These values are marked by the dotted vertical lines in Figure \ref{fig:bw_corr_hist}. The dashed lines are in the tail of the histogram with relatively high correlation values. That is, all the frequently recovered but incorrect terms have correlation values to one of the expected terms that lie in the tail of the histogram. Note however the distribution falls off into the tail very gradually and a large portion of terms reside in the tail of the distribution. This signals that overall the candidate library is redundant, leading to terms that are numerically similar to the expected terms to be identified.

Another way to test whether terms look numerically similar in the weak form is to only include the expected ideal MHD terms in the candidate library. This eliminates the possibility of incorrect terms being confused for the expected ones. This setup is similar to searching for coefficients of terms in an equation of an assumed form as in \cite{Bortz_Messenger_Dukic_2023}, but terms could be dropped via the sparsity enforcement of WSINDy. For example, using this approach with datasets below the red line of plasma $\beta$ in Figure \ref{fig:bw_recoveries} for the x-momentum equation recovers Equation \ref{eq:mom} as expected. However, for datasets above this line of plasma $\beta$, only the total pressure term $\partial_x(p^*)$ is recovered with a decreased coefficient value. Recall from the definition of $p^*$ in Equation \ref{eq:ptot} that it contains the magnetic pressure influence from the magnetic fields. This suggests that for Brio-Wu datasets with plasma $\beta$ greater than $2.25$, the magnetic field effects dominate compared to the $\partial_x(\rho v_x^2)$ term or in this limit the terms appear similar so only one is identified.

A similar analysis for the y-momentum equation shows that when all other possible terms are eliminated from the candidate library besides the expected ideal MHD terms, Equation \ref{eq:mom} is consistently recovered across $p_{left}$ and $B_0$ values. This suggests that rather than a certain effect dominating like with the x-momentum equation, here numerical similarity once in the weak form could be causing difficulty. The same issues could arise with the strong form, but the weak representation is what is used by WSINDy and therefore the focus of this discussion. Repeating this analysis with the energy equation does not aid in the recovery of Equation \ref{eq:E}, with only a subset of the expected terms being recovered.  This suggests that in some cases numerical similarity of the candidate terms in the weak form causes difficulty. In others, either the dynamics brought about by certain terms of the PDE are not significant enough compared to other effects or the theoretically expected terms are numerically similar and encompassed by a single term.

Recall that the Brio-Wu shock tube is a classic example of a Riemann problem, which when polynomial combinations of the solutions are constructed allow for redundancies. These redundancies occur in the construction of the candidate library. In regions of constant density or velocity, several equations in the system become linear conservation laws. Moreover, strong shocks (propagating through $x>0$) have constant speed, thus can be described by a linear Riemann problem. Thus the behaviors exemplified by this Riemann problem lead to the candidate library redundancies observed here.

%% file: Sedov_results.tex
\begin{figure}[t!]
    \begin{subfigure}[t]{0.5\textwidth}
        \centering
        \includegraphics[width=\textwidth]{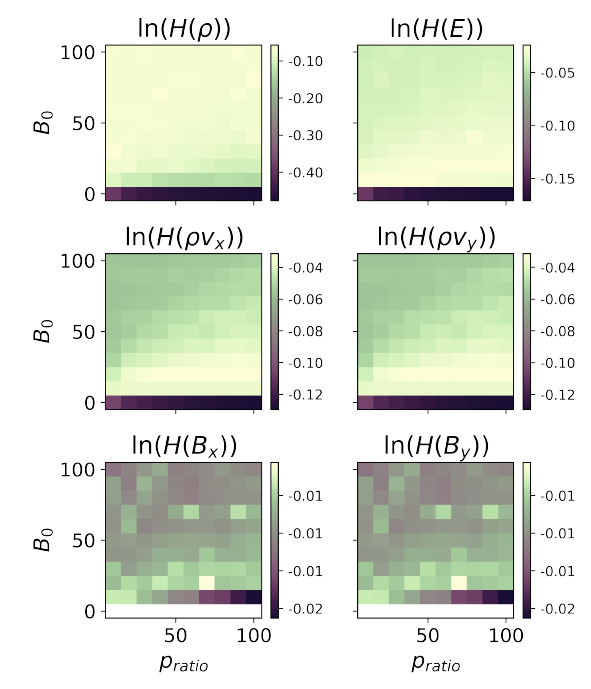}
        \caption{Information entropy calculations for density $\rho$, momentum $\rho v_x$ and $\rho v_y$, magnetic field $B_x$ and $B_y$, and energy $E$ of the magnetized sedov blast datasets. Corresponding to the initial conditions shown in Figure \ref{fig:sedov_data}, the x-axis describes an increasing pressure ratio $p_{rat}$ in the initial conditions and the y-axis represents increasing magnetic field strength $B_0$. The $B_0=0$ datasets are excluded from the lower two panels as no magnetic field is present.}
        \label{fig:sedov_ent}
    \end{subfigure}%
    ~ 
    \begin{subfigure}[t]{0.5\textwidth}
        \centering
        \includegraphics[width=\textwidth]{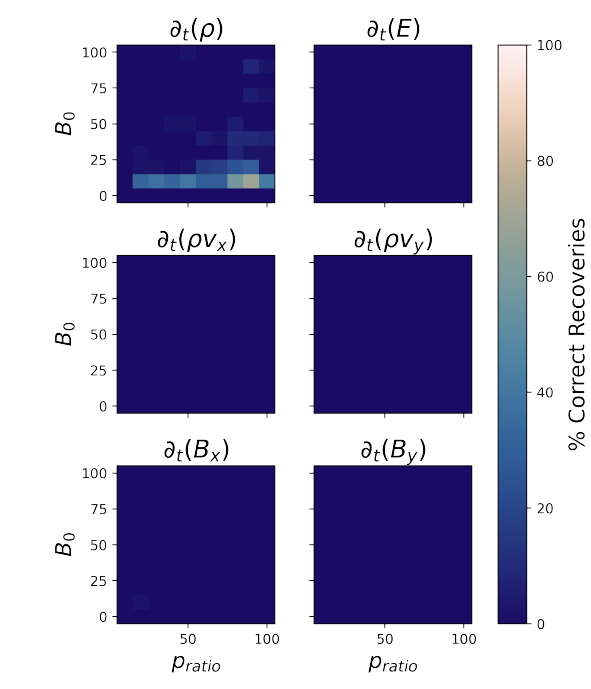}
        \caption{WSINDy recoveries of the ideal MHD equations for 40 repetitions of WSINDy per magnetized sedov blast dataset with 5\% random noise added. The continuity equation $\partial_t \rho$ is recovered correctly for high pressure ratio compared to magnetic field strength $B_0$, but the remaining equations for momentum ($\partial_t (\rho v_x)$ and $\partial_t(\rho v_y)$), magnetic fields ($\partial_t B_x$ and $\partial_t B_y$, and energy $\partial_t E$, are not ever successfully recovered.}
        \label{fig:sedov_recoveries}
    \end{subfigure}
    \caption{Shannon information entropy and WSINDy results from the magnetized sedov blast datasets. The x-axis of each plot shows the ratio of the initial circle of high pressure compared to the ambient pressure $p_{ratio}$. The y-axis represents the initial magnetic field magnitude. The overall equation recovery is poor in Figure \ref{fig:sedov_recoveries}, however, the entropy and recovery plots for density $\rho$ in the top left panel of each figure follow the same trends.}\label{fig:sedov_both}
\end{figure}

The Shannon information entropy for each ideal MHD Equation for the Magnetized Sedov blast datasets is shown in Figure \ref{fig:sedov_ent}. The entropy of density $\rho$ generally increases as you move from the bottom right to top left, while energy $E$ does the opposite. In the absence of magnetic fields ($B_0 = 0$) the entropy is low.

The results of running WSINDy over variations of the Magnetized Sedov Blast simulations are shown in Figure \ref{fig:sedov_recoveries}. While there are a few cases where the continuity equation is recovered, most of the ideal MHD equations are not recovered. Note, however, that the trend in recovery for the continuity equation echos that of entropy in the top left tile of Figure \ref{fig:sedov_ent}. \reva{Repeating this analysis without any added noise does not improve equation recovery and makes it more difficult to observe trends due to the learned equation on each dataset being deterministic. Even at these low and no noise levels, the ensemble of ideal MHD equations is not recovered from any of the Magnetized Sedov-Taylor Blast simulations.} Given the Sedov blast can be thought of as a 2D analog to the Brio-Wu shock tube, this suggests that moving to datasets with higher spatial dimensions does not necessarily improve results. The Sedov Blast is fairly symmetric, which could make recovery more challenging compared to the 1D shock tube as there are still few features but increased dimensionality. Recall the Sedov Blast is a Riemann problem with a single initial discontinuity. The Riemann behavior arising from this single discontinuity creates redundancies, making equation recovery difficult.

%% file: OT_results.tex
\begin{figure}[t!]
    \begin{subfigure}[t]{0.5\textwidth}
        \centering
        \includegraphics[width=\textwidth]{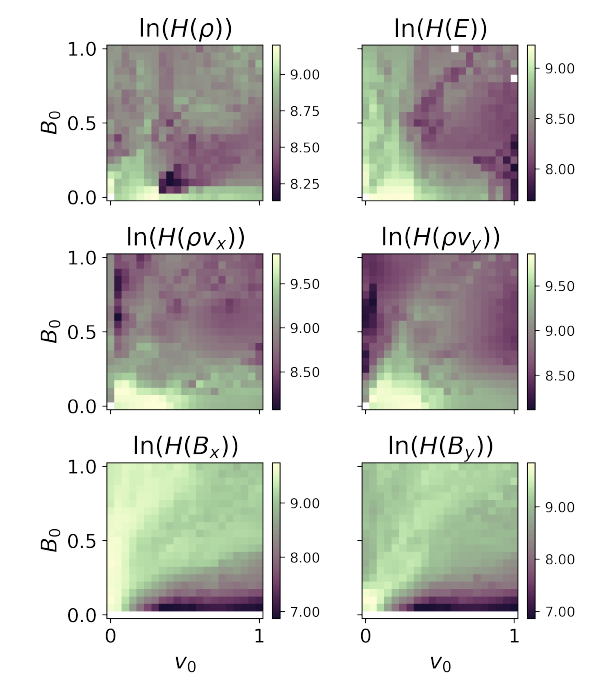}
        \caption{Information entropy calculations for density $\rho$, momentum $\rho v_x$ and $\rho v_y$, magnetic field $B_x$ and $B_y$, and energy $E$ of the Orszag Tang datasets. Corresponding to the initial conditions shown in Figure \ref{fig:ot_data}, the x-axis describes an increasing velocity $v_0$ in the initial conditions and the y-axis represents increasing magnetic field strength $B_0$. The $B_0=0$ datasets are excluded from the lower two panels as no magnetic field is present.}
        \label{fig:OT_ent}
    \end{subfigure}%
    ~ 
    \begin{subfigure}[t]{0.5\textwidth}
        \centering
        \includegraphics[width=\textwidth]{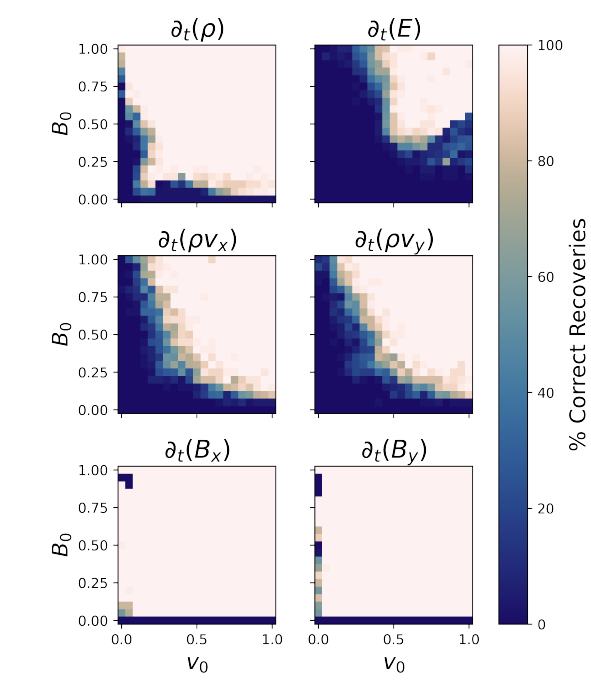}
        \caption{WSINDy recoveries of the ideal MHD equations for 40 repetitions of WSINDy per Orszag Tang dataset with 20\% random noise added. The equations for density $\partial_t \rho$ and momentum ($\partial_t (\rho v_x)$ and $\partial_t(\rho v_y)$) are consistently recovered once there is enough initial velocity $v_0$ and magnetic field $B_0$ while the magnetic field equations ($\partial_t B_x$ and $\partial_t B_y$) are recovered once a magnetic field is present. The energy equation $\partial_t E$ is the most difficult to recover, requiring more initial velocity and strong magnetic fields compared to the other equations.}
        \label{fig:OT_recoveries}
    \end{subfigure}
    \caption{Shannon information entropy and WSINDy results from the Orszag-Tang vortex datasets. The x-axis of each plot shows the initial velocity magnitude while the y-axis represents the initial magnetic field magnitude. Overall equation recovery is successful in Figure \ref{fig:OT_recoveries} once there is enough initial velocity and magnetic field. There is not a strong correlation between trends in the entropy plots and their corresponding equation recovery plots.}\label{fig:OT_both}
\end{figure}

Figure \ref{fig:OT_ent} shows the information entropy calculated per equation of the Orszag-Tang datasets. The entropy generally decreases as the magnetic field and velocity magnitudes increase. The main feature in the entropy plots is that entropy is high when the initial velocity $v_0$ and magnetic fields $B_0$ are low. This corresponds to simulations where the dynamics do not pass through the periodic boundary conditions more than once. These simulations have more coarse scale interactions. That is the entropy is high when there are fewer interactions in the fields as this leads to a more even distribution of states when $P(X)$ is calculated as in Equation \ref{eq:ent}.

An analysis of the ideal MHD equation recoveries was completed on the family of Orszag-Tang datasets and is shown in Figure \ref{fig:OT_recoveries}. In general, once the initial velocity and/or magnetic field magnitudes are large, all of the ideal MHD equations are recovered. There is no divider through the middle of the domain and divisions between more and less consistent equation recoveries do not depend on the approximate plasma $\beta$ of the system. The energy equation still has the lowest overall recovery rate as it requires higher velocities and magnetic field magnitudes before the PDE is consistently recovered when compared to the other equations for this system. The overall recovery rate for the equations in decreasing order is density, magnetic field, momentum, then energy. 

%% file: SC_results.tex
\begin{figure}[t!]
    \begin{subfigure}[t]{0.5\textwidth}
        \centering
        \includegraphics[width=\textwidth]{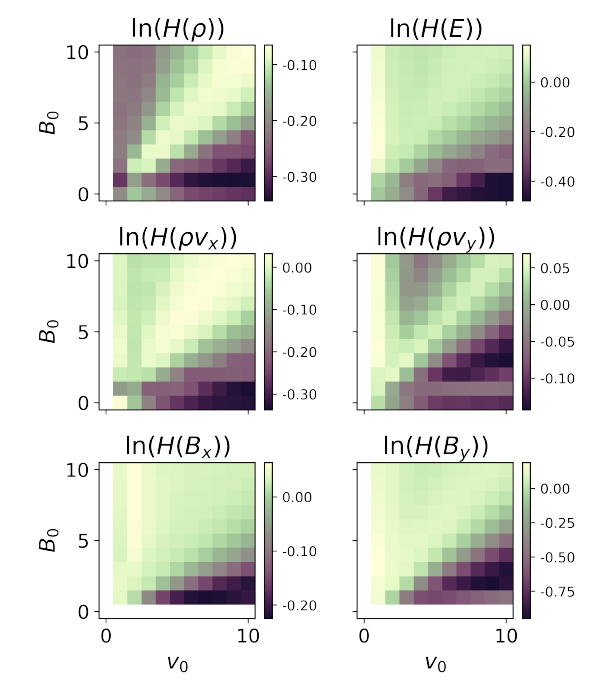}
        \caption{Entropy calculations for each equation of shock cloud datasets with $\rho_{in} = 5$. The entropy is calculated per equation of the ideal MHD equations using the data of which a subset is used to construct the vector $b$ in Equation \ref{eq:weak}. The six quantities considered are thus density $\rho$, momentum $\rho v_x$ and $\rho v_y$, magnetic fields $B_x$ and $B_y$, and energy $E$. The $v_0=0$ datasets are excluded on all plots as the dataset is constant. Similarly the $B_0=0$ datasets are excluded in the lower two panels as there is no magnetic field.}
        \label{fig:SBdens_ent}
    \end{subfigure}%
    ~ 
    \begin{subfigure}[t]{0.5\textwidth}
        \centering
        \includegraphics[width=\textwidth]{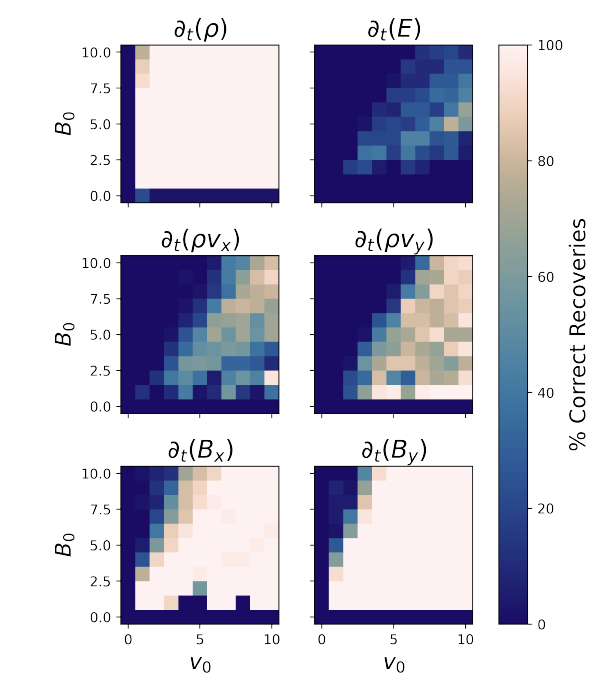}
        \caption{Ideal MHD equation recoveries for 40 repetitions of WSINDy per shock cloud data set with 1\% random noise added. $\rho_{in} = 5$, $\rho_{out} = 1$. The density equation $\rho$ is consistently recovered while the momentum $\rho v_x$ and $\rho v_y$ and magnetic field equations $B_x$ and $B_y$ have a linear relation between magnetic field strength $B_0$ and initial velocity $v_0$ above which the equations are difficult to recover. Energy $E$ also has a linear relation, however is overall more difficult to recover.}
        \label{fig:SBdens_recoveries}
    \end{subfigure}
    \caption{Shannon information entropy and WSINDy results from the shock cloud datasets. The x-axis of each plot shows the initial velocity magnitude $v_0$ and the y-axis represents the initial magnetic field magnitude $B_0$. The overall equation recovery is fairly successful in Figure \ref{fig:SBdens_recoveries}. Some equations are more consistently recovered than others, but all shared linear relationships between $v_0$ and $B_0$ can be seen between the entropy and recovery plots.}\label{fig:SC_both}
\end{figure}

Figure \ref{fig:SBdens_ent} shows the calculated entropy values for each of the shock cloud datasets. The entropy values for each equation are divided into regions by linear relationships between the initial velocity $v_0$ and the magnetic field magnitude $B_0$. While the location of these divisions differes slightly between the equations, generally the entropy is lowest in the lower right corner of each plot in Figure \ref{fig:SBdens_ent}. This corresponds to when the velocity magnitude is large compared to magnetic field magnitude. Some data points are excluded to account for when no dynamics are present. For example, when $v_0 = 0$ the velocity front never interacts with the over dense cloud so these points are excluded in all the plots. Similarly $B_0 = 0$ is excluded on the magnetic field plots.

Figure \ref{fig:SBdens_recoveries} shows the results of WSINDy in recovering the ideal MHD equations from the shock cloud datasets previously described using $\rho_{in} = 5$ and $\rho_{out} = 1$. The continuity equation is recovered as long as there is a magnetic field and an initial velocity front (i.e. $v_0 > 0$ and $B_0 > 0$). The momentum and energy equations can also be recovered, but are much less consistent than the other equations. The energy equation, as with the other datasets considered, remains the most difficult to recover. Notice also the linear relationship between velocity and magnetic field that divides regions of more and less successful equation recovery. While not directly one-to-one with the entropy plots in Figure \ref{fig:SBdens_ent}, a clear relationship can be observed. Each equation has a different line dividing more and less successful equation recovery, but these dividers are generally a linear relationship between $v_0$ and $B_0$.

While not shown, it is worth noting that decreasing the cloud density (decreasing the contrast in cloud density relative to the ambient medium) significantly worsens these results as there are regions of parameter space where the continuity equation and the magnetic field equations are recovered, however the momentum and energy equations are unsuccessful.

\begin{figure}[t!]
    \centering
    \begin{subfigure}[t]{0.5\textwidth}
        \centering
        \includegraphics[width=\linewidth]{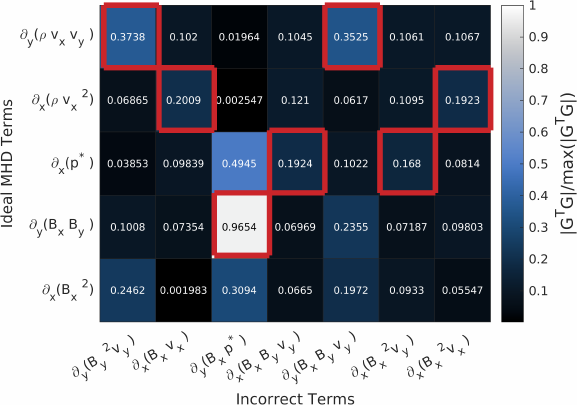}
        \caption{Example of correlation of most frequently identified unexpected terms to expected ideal MHD terms. The terms along the x-axis are chosen by identifying the most frequently recovered incorrect terms in this region from the same data displayed in Figure \ref{fig:SBdens_recoveries}. The values boxed in red highlight which of the ideal MHD terms the incorrect terms are most similar to. These also correspond to the vertical dashed lined in panel (b).}\label{fig:SBdens_corr_mat}
    \end{subfigure}%
    ~ 
    \begin{subfigure}[t]{0.5\textwidth}
        \centering
        \includegraphics[width=\linewidth]{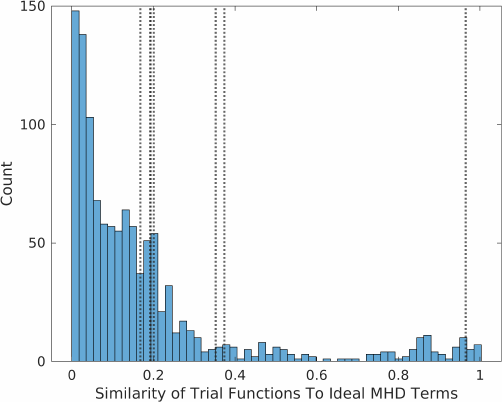}
        \caption{Histogram of correlation of all candidate library terms when compared to the expected terms for this equation. The dashed lines mark the max correlation of the incorrect terms in the x-axis of Figure \ref{fig:SBdens_corr_mat} compared to the expected terms. The vertical dashed lines show where the red boxed similarity values in panel (a) fall in the histogram.}\label{fig:SBdens_corr_hist}
    \end{subfigure}
    \caption{Analysis of correlation between candidate library terms compared to the expected ideal MHD terms for $\partial_t(\rho v_x)$ of the shock cloud dataset. These are examples from looking at the shock cloud dataset with $v_0 = 5$ and $B_0 = 8$, falling in the region of poor recovery for this equation in Figure \ref{fig:SBdens_recoveries}.}\label{fig:SBdens_corr}
\end{figure}

There is a strip of datasets in the domain where the momentum equation is not recovered, but the magnetic field equations are. Looking at one of these datasets ($v_0 = 5$ and $B_0 = 8$), we can see a similar trend as with Brio-Wu where the most commonly recovered incorrect terms look similar enough to an expected term that they could be numerically confused by WSINDy as shown in Figure \ref{fig:SBdens_corr_mat}. Looking at the overall distribution of correlations of candidate library terms to the expected terms in Figure \ref{fig:SBdens_corr_hist}, the histogram has a long tail. This suggests overall the terms are fairly unique. The threshold for incorrect terms is relatively low (as low as $0.17$), however the bulk of terms, as shown in Figure \ref{fig:SBdens_corr_hist}, lie below this value as the histogram has a long tail. Thus the commonly-recovered but incorrect terms are in the tail of this histogram, meaning while the bulk of the candidate library is easily distinguished from the expected terms, there are enough similar terms that equation recovery is difficult.

%% file: Discussion.tex

The results presented in Section \ref{sec:results} demonstrate that not all datasets are equivalent for our purposes; in other words, certain datasets are easier to discern governing dynamics from. Problems involving a single prominent shock or feature, like with the Brio-Wu shock tube or Magnetized Sedov Blast, are dominated by those effects so the full description of the underlying dynamics cannot be robustly recovered. In some cases this is due to the fact that in these types of setups there is redundant behavior and therefore numerically similar terms in the candidate library. For example in looking at the similarities between terms for Brio-Wu in Figure \ref{fig:bw_corr_mat}, $\partial_x(\rho v_x)$ looks nearly identical to the desired term $\partial_x(\rho v_x v_y)$, with the similarity measure between these two terms being one in Figure \ref{fig:bw_corr_mat}. In some cases even expected terms could look equivalent meaning all expected terms cannot be recovered.

Systems described by a larger number of interactions between components, like the Orszag-Tang Vortex and shock cloud problems, do not have the same redundancies. These types of problems have a much smaller region where the equations are not recovered, and their candidate library terms are more unique. For example, in the histogram of correlation between candidate library terms to expected terms for the shock cloud datasets in Figure \ref{fig:SBdens_corr_hist} the majority of the terms have correlations near zero, meaning they are more unique and easily eliminated by WSINDy compared to the expected ideal MHD terms. Compared to the distribution of correlations for the Brio-Wu shock tube in Figure \ref{fig:bw_corr_hist}, the tail of the histogram is very long, meaning the majority of terms are easily distinguishable from the ideal MHD terms. The correlation histogram for the Orszag-Tang candidate library is even more extreme than the shock cloud, having a large spike near zero with a long tail spanning the remainder of the domain. The shape of these histograms mean these problems have candidate libraries with more unique and therefore distinguishable terms. The datasets with more complex interactions lead to correlation distributions with longer tails and overall more unique candidate libraries. The complexity in the data that pertains to a specific governing equation, rather than simply the dimensionality of the data, dictates what PDE terms can be identified. \revc{Model identification on these datasets is also more robust to noise compared to the other datasets}.

The shock cloud problem could be thought of as intermediate between the datasets with a single initial feature (Brio-Wu and Sedov) and Orszag-Tang which contains many interactions. The shock cloud equation recoveries in Figure \ref{fig:SBdens_recoveries} are similar to those of Brio-Wu in that for many of the equations there is a clear division through the middle of the domain where the ideal MHD equations for momentum, magnetic fields, and energy are recovered on one side of the line but not the other. Recall using a limited library with Brio-Wu eliminated these issues for some simulations, but not others. This suggests that some of this difficulty in recovery could be due to numerically similar terms being present in the candidate library. In the unremedied cases the theoretically known terms in the PDE may be numerically similar due to the Riemann nature of the problem, making the known terms difficult to distinguish. The Magnetized Sedov Blast and Brio-Wu are both Riemann problems, where much of the dynamics are based at the same physical locations, which could worsen this problem. The shock cloud problem contains discontinuities, however these discontinuities interact leading to more numerically dissimilar fields and therefore candidate library terms. Considering the overall fraction of the domain where the ideal MHD equations are recovered in Figure \ref{fig:SBdens_recoveries}, some similarities to the Orszag-Tang recoveries in Figure \ref{fig:OT_recoveries} can be observed. For example the continuity equation is the most consistent, followed by the magnetic field equations, then the momentum equations, and finally the energy equation.

Following this thought process then leads back to the information entropy of the system. The entropy of a system generally increases as the probability distribution flattens \cite{engelmann_towards_2022}, which in our case translates to the system being observed an equal amount in all its states. For the Brio-Wu datasets, which are defined by a few identifiable features that expand outward \cite{BW_waves}, there are regions of parameter space that lead to less variability for a given component. For example, looking at the information entropy for the Brio-Wu datasets in Figure \ref{fig:BW_ent}, below the line $\beta=2.25$ the x-momentum equation has lower entropy, however the y-momentum equation has greater entropy. Compared to the equation recoveries in Figure \ref{fig:bw_recoveries} then, for the momentum equations the regions with low entropy have improved recovery. However in comparison to the other equations, the recovery is worse in regions of low entropy. The momentum is a product of two evolved quantities, density $\rho$ and velocity $v$, so this may be a measurement of how much these two quantities interact. Thus while entropy does echo the trends in recovery for this problem, it is not directly correlated to equation recovery. Rather the information entropy seems to be informing where there are repeated behaviors in the data. These behaviors could make recovery easier, in the case of the momentum equations, or more difficult, in the case of the remaining equations. It depends on how indicative of the PDE the repeated behavior is.

As the complexity of the data increases however, entropy seems to lose its informative behavior. For example, for Orszag-Tang in Figure \ref{fig:OT_ent} the most significant feature in the entropy calculations is high entropy for lower initial velocity and magnetic field amplitudes. In the simulation data this corresponds to where behavior has not propogated around the periodic boundary conditions and interacted at the center of the domain. For the magnetized shock cloud the trends limiting recovery can be seen in the information entropy in Figure \ref{fig:SBdens_ent}, but not as strongly as for Brio-Wu nor as weakly as for Orszag-Tang. The information entropy of the system becomes less descriptive as the interactions in the data become more complex and there is more dimensional coupling. The more complex datasets display stronger behavior brought about by interactions between components and derivatives in different dimensions. At the same time, this increase in interaction complexity aids equation recovery. Information entropy is a qualitatively informative metric for problems with strong limiting behavior that is hard to recover equations from, but not descriptive for problems where equation recovery is easier. Thus information entropy is most helpful when studying problems where equation recovery is difficult due to the data containing a single dominant behavior, like a strong shock or Riemann problem.

These results may also relate to the complexity of the PDEs being studied. The ideal MHD equations considered here are much more complex than those classically considered when testing sparse regression data-driven methods. The equations usually used as test cases do not have nearly as many components, nor do they have as many equations required to describe the system evolution when compared to the ideal MHD equations (equations \ref{eq:cont} through \ref{eq:E}). Further, the ideal MHD equations have more complex interactions between dimensions brought about by the cross products and gradients. The simulations with a few significant features, like Brio-Wu and the Sedov blast, may lack the data complexity needed to recover these more complex equations due to the redundancies in terms. So one could think of these challenges in recovery as a relationship between the complexity of interactions in the data related to the level of dimensional coupling in the desired PDEs. Further, in some cases a simplified or linearized form of the equations may be recovered, which by our analysis would be counted as incorrect. Linearized forms tend to occur for datasets with smaller initial magnitudes in magnetic fields, velocity, or pressure.

%% file: Conclusions.tex
Our results show that recovering the ideal MHD equations from simulation data can be challenging depending on the type of features a particular simulation demonstrates. Simulations with only a few descriptive features exist in a low information limit where specifics of the problem setup may not bring about interactions descriptive of certain PDE terms. In this low data limit, information entropy is able to pickup on the trends in parameters that lead to simulations lacking robust information for full equation recovery. While it is not a perfect metric \revc{as other data or method properties beyond low information may impact equation recovery,} it does point in the right direction. Equations that involve dimensional coupling require significant enough interactions in the data to recover the full PDE. \revc{The energy equation is the most challenging to recover across all datasets, but further work is needed to conclude whether this is a problem unique to ideal MHD or a more wide spread finding.} Riemann problems lead to redundancies in the data and constructed candidate libraries that make equation recovery especially difficult. Riemann problems for ideal MHD can be seen as degenerate from the perspective of model selection, or lacking sufficient information, unless sufficiently large jumps in density, magnetic field, or pressure are present to trigger nonlinear effects. Even for problems that lead to more descriptive behavior, like Orszag-Tang or the Shock Cloud problem, there is a low data limit where the equations can no longer be recovered. This echos results previously shown in \cite{wsindy_pde} where for high enough noise levels the Navier-Stokes equation simplified to the Euler equation. The required data complexity observed here can also be influenced by the noise level, which can obscure important behavior.

Given this need for complexity of interactions in the data, WSINDy and methods like it may perform better when trying to move from a high fidelity simulation like particle-in-cell (PIC) to a simplified fluid model. It has already been shown that WSINDy like methods can recover fluid equations from PIC data \cite{alves_data-driven_2020}. They recovered the MHD equations from a PIC simulation of a propagating shock, but the PIC data used leads to smaller scale perturbations around the actual shock front. From the results presented here, these smaller scale interactions increase the complexity in the data, making equation recovery easier. Thus using WSINDy on a system like the DPF, which starts in a regime requiring PIC simulations, may be highly successful in discovering simplified fluid models throughout its evolution. Another way to reliably use this method could be to assume an equation form and only search for unknown coefficients. In the cases presented here, when the correct PDE form is recovered the coefficients are very accurately recovered. Generally the coefficients have at most 15\% error for the more challenging datasets and much less for the datasets with more complex interactions. A variation of WSINDy that only searched for coefficients already exists, so using these plasma simulations that are more challenging to work with than the PDEs usually used could be a compelling test \cite{Bortz_Messenger_Dukic_2023}.

We would hope to find a metric to quantify this relation between dimensional coupling and complexity in the desired PDE compared to what is observed in the data. Entropy is helpful for the less complex simulations, but is not a global descriptor relating back to the underlying PDE. A more sophisticated metric is needed to globally quantify dataset complexity. A metric worth considering could be the correlation dimension of the problem, which estimates the minimum number of dimensions a problem can be represented in \cite{Michel_Flandrin_1996}. If the problem exists in two spatial dimensions, but can be represented on a lower dimensional manifold, it may not be possible to recover the full set of PDEs from that particular dataset. \reva{Instead, the sparsity enforcing component of WSINDy may be trying to identify dynamics in this lower dimension.} \reva{Results in other research have suggested a similar possibility \mbox{\cite{Kaptanoglu_Zhang_Nicolaou_Fasel_Brunton_2023}}}.

Another helpful metric would be one that can eliminate redundant terms in the candidate library. There are methods like relieff that rank features predictive potential, but brief tests of this method show that it struggles even more than WSINDy on the ideal MHD problems studied here \cite{ReliefF_Kononeko}. Ensemble methods have also been helpful with this on other problems, but would be interesting to study on these challenging plasma problems \cite{Fasel}. Alternatively, the redundancy analysis presented here is that of the trial functions once convolved with the test functions. The selection of query points and support widths is then an important component of the algorithm as this determines what regions of the spatiotemporal domain contribute to the weak representation. Other work has demonstrated that the ratio of the support width to system scales strongly influences whether the proper functions are recovered \cite{Messenger_Burby_Bortz_2024}. Thus it may be that the ideal MHD problems studied here highlight a weakness in the automatic hyperparameter selection process that is dependent on conditions of the particular dataset instead of the underlying governing equations.